\documentclass[twocolumn]{bmcart}

\usepackage{lineno}
\usepackage{cite}

\usepackage{graphicx}
\usepackage{subfigure}
\usepackage{amsmath}

\begin{document}

\begin{frontmatter}
\begin{fmbox}
\dochead{Research}

\title{Characterization of a scintillating lithium glass ultra-cold neutron detector }



\author[
  addressref={aWpg},
  corref={aWpg},
  email={bl.jamieson@uwinnipeg.ca}
]{\inits{B}\fnm{Blair}~\snm{Jamieson}}
\author[
  addressref={aWpg}
]{\inits{L.A.}\fnm{Lori A}~\snm{Rebenitsch}}
\author[
  addressref={aWpg}
]{\inits{S.}\fnm{Sean}~\snm{Hansen-Romu}}
\author[
  addressref={aPSI}
]{\inits{B.}\fnm{Bernhard}~\snm{Lauss}}
\author[
  addressref={aTRIUMF,aWpg}
]{\inits{T.}\fnm{Thomas}~\snm{Lindner}}
\author[
  addressref={aWpg}
]{\inits{R.}\fnm{Russ}~\snm{Mammei}}
\author[
  addressref={aWpg}
]{\inits{J.W.}\fnm{Jeff W}~\snm{Martin}}
\author[
  addressref={aRCNP,aTRIUMF}
]{\inits{E.}\fnm{Edgard}~\snm{Pierre}}


\address[id=aWpg]{
  \orgname{Department of Physics, University of Winnipeg}, 
  \street{515  Portage Avenue}, 
  \city{Winnipeg},
  \cny{Canada}
}

\address[id=aPSI]{
  \orgname{Laboratory for Particle Physics, Paul Scherrer  Institute}, 
  \street{CH5232},
  \city{Villigen PSI}, 
  \cny{Switzerland}
}

\address[id=aTRIUMF]{
  \orgname{TRIUMF}, 
  \street{4004 Wesbrook Mall}, 
  \city{Vancouver}, 
  \cny{Canada}
} 

\address[id=aRCNP]{
  \orgname{Research Centre for Nuclear Physics, Osaka University},
  \city{Osaka},
  \cny{Japan}
}

\end{fmbox}

\begin{abstractbox}

\begin{abstract}

A $^{6}$Li glass based scintillation detector developed for the TRIUMF
neutron electric dipole moment experiment was characterized using the
ultra-cold neutron source at the Paul Scherrer Institute (PSI).  The
data acquisition system for this detector was demonstrated to
perform well at rejecting backgrounds. An estimate of the absolute
efficiency of background rejection of $99.7\pm0.1$\% is made.  For
variable ultra-cold neutron rate (varying from $<$~1~kHz to
approx.\ 100~kHz per channel) and background rate seen at the Paul
Scherrer Institute, we estimate that the absolute detector efficiency
is $89.7^{+1.3}_{-1.9}$\%.  Finally a comparison with a commercial
Cascade detector was performed for a specific setup at the West-2
beamline of the ultra-cold neutron source at PSI.

\end{abstract}


\end{abstractbox}

\end{frontmatter}

\section{Introduction}

Determining the neutron Electric Dipole Moment (nEDM) limits theories
beyond the Standard Model~\cite{pospelov}.  Ultra-Cold Neutrons (UCN)
provide a good means to search for a nEDM.  As a result, there are
various nEDM experiments around the world utilizing UCN that are
either running or being
planned~\cite{fillipone,PSI,Gatchina,APSerebov,KKirch,CABaker,YMasuda,IAltarev,RGolub,SNS}.
Measurements are limited mainly by UCN statistics.  Increasing the
efficiency of the detection system is therefore important.

The UCN source at the Research Centre for Nuclear Physics (RCNP) in
Osaka successfully demonstrated UCN production in super-fluid helium
and extraction through cold windows in 2013~\cite{kawasaki2014}.  This
source is in the process of being moved from RCNP to TRIUMF, in
Vancouver, over the coming year where a new UCN facility is being
prepared.  A neutron Electric Dipole Moment (nEDM) experiment is
planned as the first experiment after the source will be installed at
TRIUMF~\cite{jeffucn}.

A UCN detector using $^6$Li glass has been designed and built for the
nEDM experiment.  This detector must fulfill several performance
requirements. The first requirement is to be able to count UCN with a
stability of 0.03\% ($1/\sqrt{10^7}$) over the hour required to measure
a few Ramsey cycles in an nEDM experiment.  A second requirement is to
dependably count UCN at high rates ( $>$~1~MHz).  Finally the
detector's sensitivity to backgrounds, needs to be well known or
measurable during periods without UCN.  In order to determine the
detector's performance, the detector has been bench-marked against a
Cascade UCN detector\footnote{CD-T Technology, Hans-Bunte Strasse
  8-10, 69123 Heidelberg, Germany} using the UCN source at the Paul
Scherrer Institute (PSI) in Switzerland~\cite{BLaussAIP2012,
  BLaussHFI2012, BLaussPP2014}.

This paper describes the $^{6}$Li based scintillation detector in
Section~\ref{sec:overview}.  One goal of the tests is to estimate the
overall detection efficiency and background rejection capabilities of
the $^{6}$Li detector.  A simulation of the UCN detection and
background detection was prepared, as described in
Section~\ref{sec:sim}.  To get an estimate of the absolute detector
efficiency, described in Section~\ref{sec:eff}, we have taken account
of the neutron selection cut efficiency, along with estimates of the
geometrical acceptance, and neutrons lost to the lithium depleted
layer of glass on top of the detector.  A comparison of the detector
measurement to a Cascade UCN detector is described in
Section~\ref{sec:relative}.

\section{Overview of Detector Technology}\label{sec:overview}


\subsection{$^6$Li Scintillating Glass Detector}

The scintillating glass is doped with $^6$Li, which has a high neutron
capture cross-section of order $10^5$~bn at UCN energies.  The charged
particles in the reaction:
\begin{equation}
^6{\rm Li} + n \rightarrow \alpha (2.05\ {\tt MeV}) + {\rm t} (2.73\ {\tt MeV})
\end{equation}
are detected. 

In order to reduce the effect of an $\alpha$ or triton escaping the
glass, two optically-bonded pieces of scintillating glass are used.
This type of scintillating stack detector was pioneered by the group
at LPC Caen~\cite{ban, ban05, afach, gban2016}. The upper layer is
$60$~$\mu$m thick depleted $^6$Li glass (GS30), and the lower layer is
$120$~$\mu$m thick doped $^6$Li glass (GS20), which allows the
resultant particles to deposit their full energy within the
scintillating glass\footnote{GS20 and GS30 were purchased from Applied
  Scintillation Technologies, now Scintacor, 8 Roydonbury Industrial
  Estate, Horsecroft Road, Harlow, CM19 5BZ, United Kingdom}.  The
$^{6}$Li content and density of these scintillators is summarized in
Table~\ref{tab:LiGlass}.

 \begin{table}
\begin{center}
\caption{Properties of the glass scintillators}\label{tab:LiGlass}
\begin{tabular}{  c  c  c }
  \hline
  Scintillator             &  GS20                   & GS30                  \\
                           &  $^6$Li enriched         & $^6$Li depleted        \\
  \hline
  Total Li content (\%)       & 6.6                  & 6.6                   \\
  $^6$Li fraction (\%)     & 95                   & 0.01                   \\
  $^6$Li density (cm$^{-3}$)\cite{spowart1976}   & 1.716$\times$10$^{22}$  &   1.806$\times$10$^{18}$  \\
  \hline
\end{tabular}
\end{center}
\end{table}

Optical contacting of the two layers was performed by Thales-Seso in
France, and a method of checking the doped side of the glass was
developed at the University of Winnipeg~\cite{jamieson}.  The
scintillation light is then guided via ultra-violet transmitting
acrylic light-guide to its corresponding photomultiplier tube outside
the detector vacuum region.  Each of the nine tiles of scintillating
glass's light is detected by a Hamamatsu R7600U Photomultiplier Tube
(PMT).  The scintillation following neutron capture gives a fast event
signal with rise time of 6~ns and a fall time of about
55~ns~\cite{atkinson1987,brollo1990, sakamoto1990}.  There is also a
slower decaying light component up to 2~$\mu$s.

The detector design is similar to the detector developed for the nEDM
experiment at PSI, and also employed at PSI for UCN
monitoring~\cite{gban2016,LGoeltl}.  These detectors have some
sensitivity to gamma-ray and thermal neutron backgrounds.  Background
contamination largely due to gamma-ray interactions in the
light-guides is discussed further in the paper in
Sections~\ref{sec:digisim} and~\ref{sec:bgreject}.

Making the scintillating Li glass as thin as possible reduces this
sensitivity to both thermal neutron captures and to $\gamma$-ray
scintillation backgrounds.  The mean range of the $\alpha$ is
5.3~$\mu$m and the mean range of the triton is 34.7~$\mu$m, meaning
that thinner than about $50$~$\mu$m could also result in an efficiency
loss as the charged particles produced in the neutron capture escape
the glass before stopping.  In addition, the gamma ray interactions in
the light-guides can be rejected by Pulse-Shape Discrimination (PSD)
since these signals do not have a slow decaying component and are
therefore shorter (FWHM approx. 20~ns) than the scintillation signal
from the lithium glass.

\begin{figure}[!htpb] 
\includegraphics[width=0.45\textwidth]{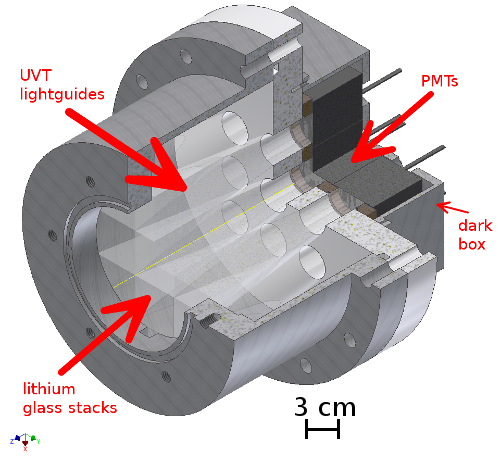} 
\includegraphics[width=0.45\textwidth]{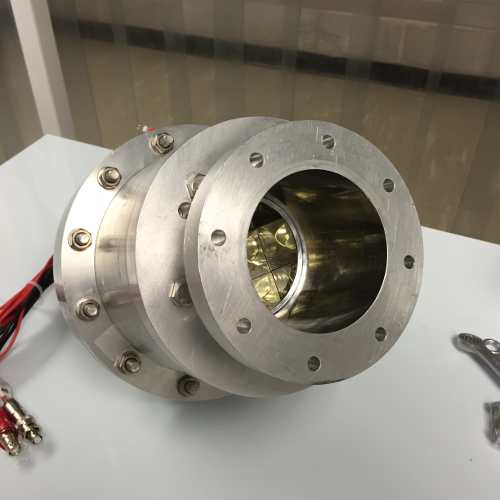} 
\caption{\label{fig:detdesigned} Three dimensional drawing of the UCN
  detector and enclosure (top), and a photo of the detector (bottom).
  The detector enclosure is made of Al, and the rim of the adapter
  flange which UCN can hit is coated with $1$~$\mu$m Ni by thermal
  evaporation (colour online).}
\end{figure}

In order to handle UCN rates up to $\sim$1~MHz, the $^6$Li detector
face is segmented into 9 tiles.  This reduces pile-up.  A
photograph and a drawing of the detector are show in
Fig.~\ref{fig:detdesigned}.  Details about the detector readout are
presented in Section~\ref{sec:signaltreat}.

The detector enclosure was machined from Al, and an adapter flange
which has a rim which UCN can hit was coated with $1$~$\mu$m
of natural abundance Ni.  The ${^6}$Li glass tile side lengths are
29~mm, and the opening on the adapter flange is 81~mm.

\subsection{$^6$Li Detector Signal Treatment}\label{sec:signaltreat}

Signals from the PMTs are amplified by a Phillips 775 octal $10\times$
preamplifier and read out directly by 8-channel CAEN V1720 digitizers.

The CAEN V1720 has a PSD firmware that triggers on pulses below a
certain threshold independently for each channel.  The digitizer
samples the waveforms every 4~ns, and for each sample digitizes the
voltage on a $2$~V scale into an ADC value between 0 and 4096.  Each
channel of the digitizer triggers when a pulse goes some number of ADC
counts below a pedestal value.  The digitizer threshold for triggering
was set at 250 ADC ($\sim -125$~mV).

The digitizer can be run with a fixed pedestal, or a pedestal taken
from an average over the last 32 samples (128 ns).  This self
calculated pedestal is called a baseline in the digitizer
documentation, and once a trigger happens, the baseline is held
constant until the end of a specified gate time.  The self-calculated
baseline was used for the detector tests described in this paper.

For each trigger, the PSD firmware calculates the sum of the signal
below the baseline starting from the trigger time for a short gate
width, $t_s=40$~ns, and for a long gate width, $t_L=200$~ns.  The
short gate time has been chosen to contain all of the charge for gamma
ray interactions in the light-guides.  The ADC sum below the baseline
within the short gate is called, $Q_S$, and the sum within the long
gate is called, $Q_L$.  The charge $Q_L$ contains the total charge
deposit for neutron capture events.  The PSD value is also calculated,
and defined as:
\begin{equation}
{\tt PSD} = \frac{ (Q_L - Q_S)} { Q_L }.
\end{equation}
After each trigger, the digitizer channel is busy for a 150~ns
dead-time.  A cut on $Q_L$ and PSD provides a rejection of gamma
interactions in the light-guides as described in
Section~\ref{sec:digisim}.  The PSD variable has been chosen to cancel
out channel to channel variations in gain by making a ratio of
charges.  Also the PSD is sensitive to the difference in shape of the
scintillation signals, and the signals from interactions in the
light-guides.

The digitizer firmware stores only the $Q_S$, $Q_L$, PSD, baseline,
and trigger time for each pulse, thereby permitting the V1720 to
handle data rates up to $\sim$2~MHz without saturating the data path.
The first seven tiles of the detector are read out on one digitizer,
and the last two tiles are read out on a second digitizer.  Each of
the digitizers was connected by an independent optical fibre to a CAEN
A3818 PCI Express card on the Data Acquisition (DAQ) computer for
readout, allowing data rates up to 85~MB/s.

The time-stamp provided by the digitizer is a clock cycle count in
4~ns ticks up to 17~seconds.  To help keep track of the time-stamp
wrap-around, and check the digitizer synchronization, a $1$~Hz pulser
was fed into the last channel of each of the two digitizers.  In
addition a signal from the PSI proton beam timing was sent into one of
the channels of the detector to be used to determine the times when
the proton beam arrived.

The DAQ software used the MIDAS system that is commonly used at PSI
and TRIUMF.  A MIDAS frontend for the CAEN V1720 was written to
collect the PSD data from the digitizers and save it into MIDAS banks.

\section{ Detector tests with UCN }

We used the two beamlines called ``West-1'' and ``West-2'' at the PSI
UCN source~\cite{ucnBeam}.  West-2 offers the distinct feature that
UCN have a dropping height of minimum 120~cm before reaching the
detector.  West-1 in a horizontal configuration provides UCN with
energies starting above 54~neV given by the safety AlMg$_3$ foil in
the beamline.

\subsection{ Time distribution }

\begin{figure}[htpb] 
\begin{center} 
\includegraphics[width=0.45\textwidth]{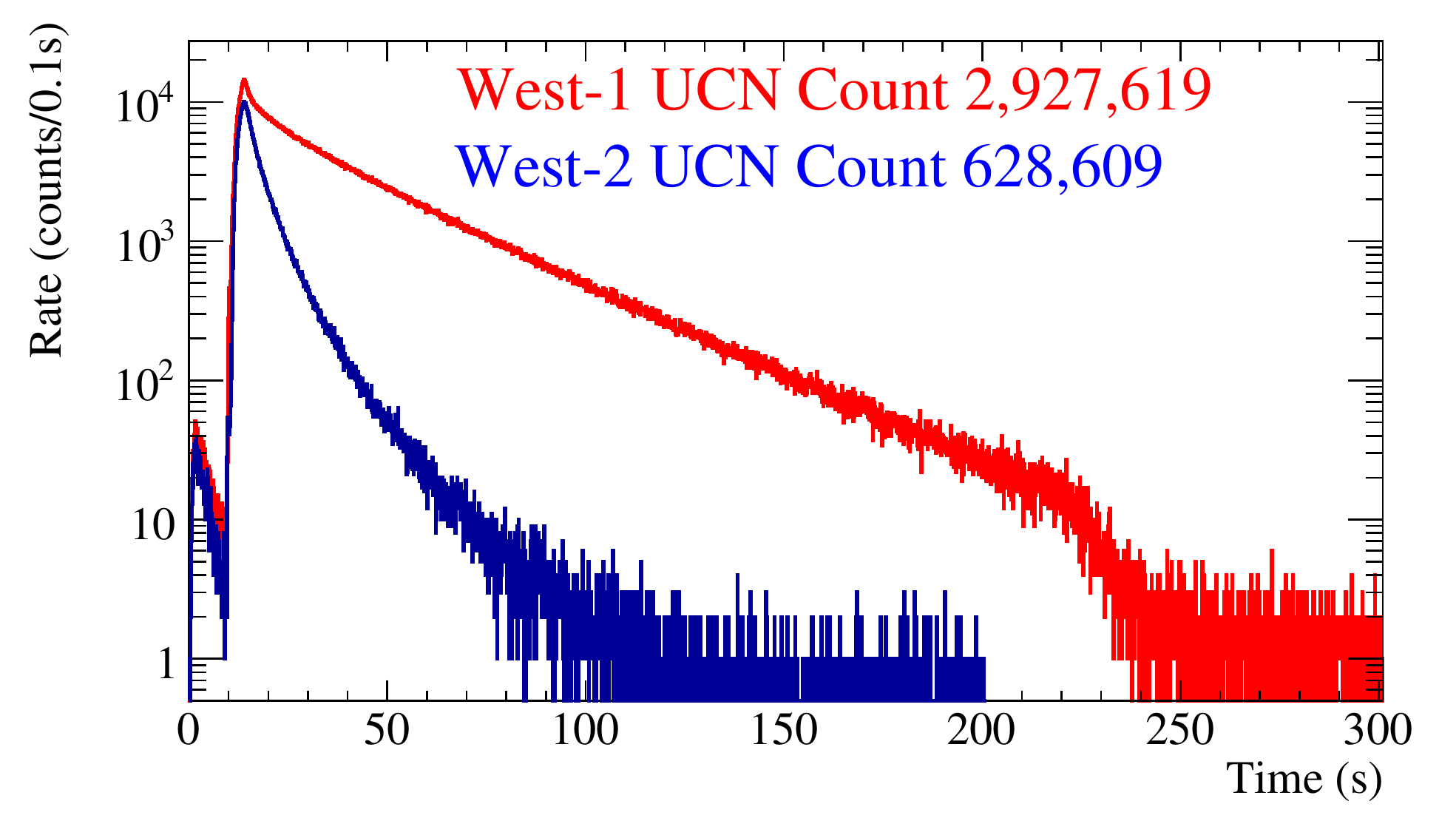} 
\caption{ UCN observed during one UCN cycle, as detected with the
  $^6$Li UCN detector on West-1 (red) and West-2 (blue).  The small
  peak before the main peak is caused by a 7~ms long pilot proton beam
  bunch used for checking the beam centering before the main beam
  bunch on the spallation target (colour online). }
\label{fig:protonCycle} 
\end{center} 
\end{figure}

During August 2015 there was a 300~s UCN cycle at PSI, where the rate
of UCN is highest right after a 3 second proton beam bunch on a
neutron spallation target.  For the next 297 seconds after the proton
beam is turned off the UCN rate falls, going from rates of tens of kHz
down to 20~Hz as seen in Fig.~\ref{fig:protonCycle}. 

UCN rates on the West-1 beamline are a factor of 10 higher than on
West-2.  The operation mode was given by the priority of the nEDM
experiment measuring at the third beam-port.  Typically, UCN were only
delivered to the West-1 beamport after 30 seconds, when the nEDM
experiment stopped its filling period.  During the UCN cycles the UCN
propagate down the beam line to the experiment area.  Typical UCN
detector rates during 300 second UCN cycles in West-1 and West-2
measured in Aug. 2015 with our scintillation detector are shown in
Fig.~\ref{fig:protonCycle}.

Note that our detector's 75~mm diameter aperture does not match the
180~mm aperture of the West-1 beam-line, and that it can only see UCN
above the Fermi potential of the scintillating glass (103.4 neV).
This means that we see relatively more UCN from the vertical source,
which has a spectrum starting at about 120~neV, matching our detector
Fermi potential, than from the softer source of the horizontal West-1
beam.  The PSI group has made measurements showing that on West-1,
about 32\% of the UCN are between 54~neV and 120~neV.  Also, the two
UCN cycles shown in the figure were taken on different days, and we
know from measurements that during that time, the UCN source saw about
a 15\% decrease in the delivered UCN intensity.  Therefore, we expect
to see less than the 10$\times$ difference in UCN rate when comparing
the distributions in Fig.~\ref{fig:protonCycle}.

\subsection{ Charge measurement }

\begin{figure}[!htpb]
\centering
\includegraphics[width=0.45\textwidth]{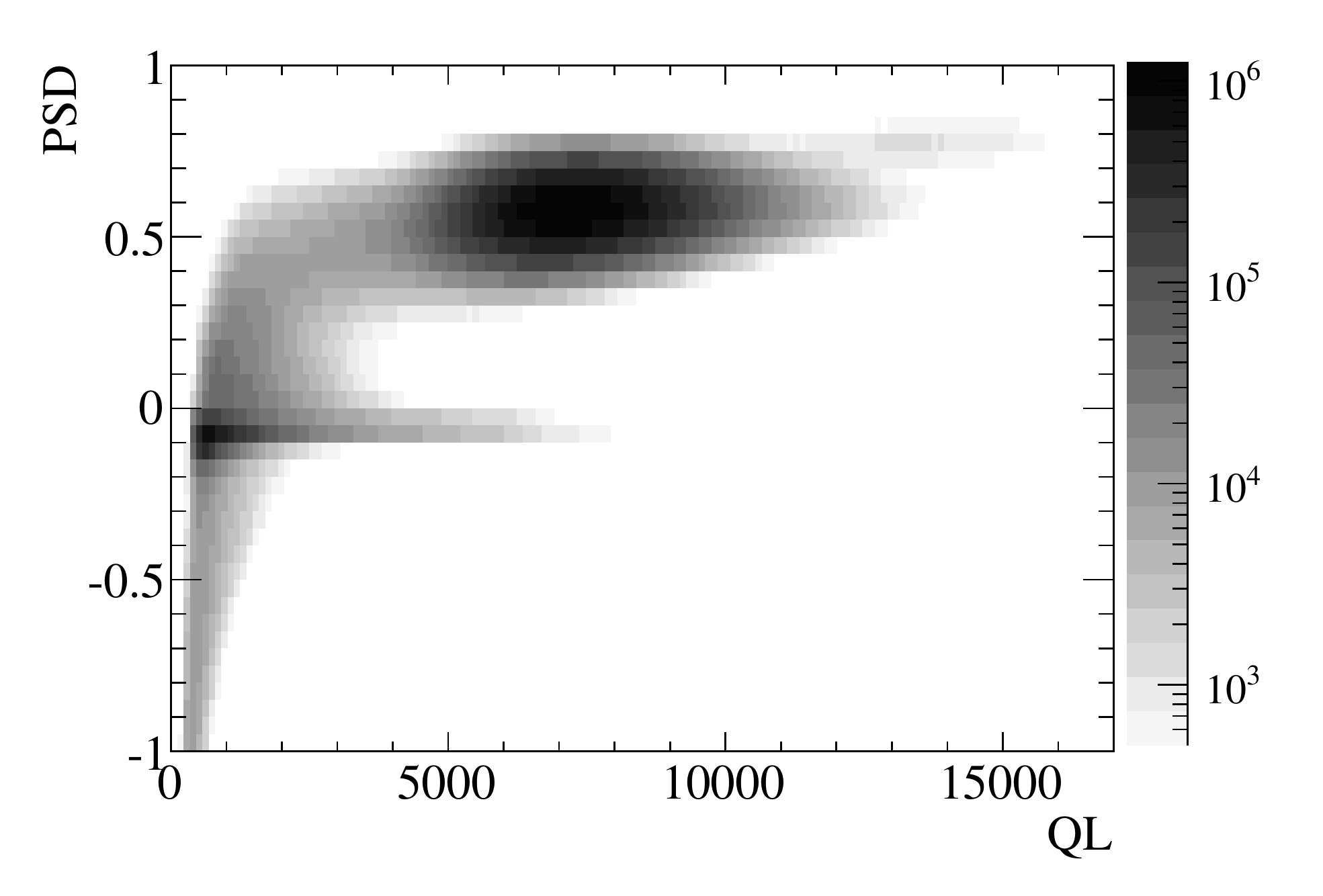}
\caption{ Event counts per bin are shown by the greyscale as a
  function of PSD and $Q_L$ from UCN data taken on the West-1
  beamline.} \label{fig:DataeventSpectra}
\end{figure}

The PSD versus $Q_L$ distribution from a UCN data run on West-1
beamline is shown in Fig.~\ref{fig:DataeventSpectra}.  The UCN signal
events are centered around a PSD of 0.5 and $Q_L \sim 5000$ to $\sim
12000$, and signals from $\gamma$-rays in the lightguides are around
$PSD \sim 0$.  The values between these are due to pile-up effects,
events right after the dead-time and late-light events.  Further
details on these effects are discussed in Section~\ref{sec:sim}. Note
that the negative PSD values come from pile-up and dead-time events.
In these events, the average baseline calculated by the digitizer
firmware is too large, resulting in the integrated $Q_L$ that it
calculated becoming smaller than the $Q_S$.  This feature of the PSD
is also seen in simulations as described in Section~\ref{sec:digisim}.

\begin{figure*}[!htpb]
\centering
\includegraphics[width=0.95\textwidth]{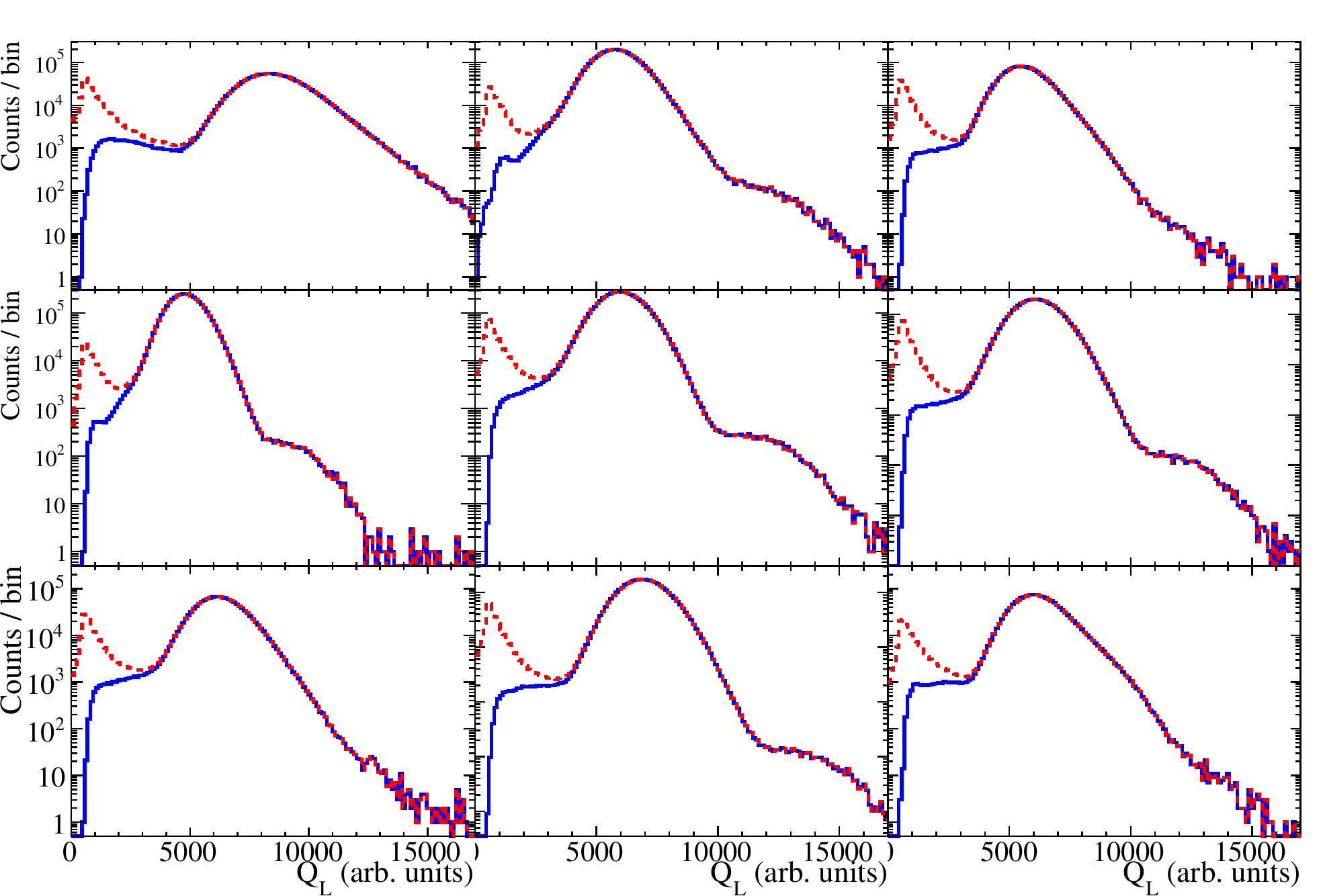}
\caption{ Event count per $Q_L$ bin from each of the nine channels of
  the $^{6}$Li scintillation detector from twelve cycles of UCN beam
  on the horizontal West-1 beamline.  The distribution without a cut
  on PSD is shown as a dashed-red line, and with a cut on PSD$>0.3$ as
  the solid-blue line (colour online).}
\label{fig:nineup}
\end{figure*}

The $Q_L$ distribution from each of the nine channels of the detector,
with a PSD cut at 0.3 to eliminate the light-guide backgrounds, is
shown in Fig.~\ref{fig:nineup}.  These data are from twelve UCN cycles
collected from the horizontal West-1 beamline.  It is clear from these
distributions that there is very little background remaining.  The
different numbers of counts seen is explained by the geometry of the
detector.  The corner square tiles are shadowed the most by the round
aperture of the detector opening, giving them the lowest count.

\subsection{ Channel to channel comparisons }\label{sec:chantochan}

In order to determine the rate stability of the detector, the ratio of
the rates $R_r$ in the outer channels to the central channel is
calculated.  This ratio of rates is compared to the ratio of area
($R_a$) of the outer channel as measured from the photograph in
Fig.~\ref{fig:detface} to the area of the central channel.

During each UCN cycle on the West-1 beamline that was used for this
comparison, the $^6$Li detector counted $\sim 3\times10^6$ neutrons
corresponding to $\pm0.06$\% statisitical uncertainty.  An overnight
run containing 114 UCN cycles (9.5 hours) was taken to assess the rate
stability.  The count for each UCN cycle in each channel, and the
ratio of counts per UCN cycle in each channel over the central channel
was plotted versus UCN cycle to assess the rate stability. Fitting the
ratio of count rates in each of the outer channels to a constant
yielded an acceptable $\chi^2$ per degree of freedom as summarized in
Table~\ref{tab:relrates}.  The difference between the relative area
and relative rate ($R_a-R_r$) is also tabulated.

We make two conclusions from these measurements.  First is that the
difference in efficiency between the channels is at most 5\%.  These
differences could be due to differences in surface contamination or
differences in the number and energy of UCN reaching the different
tiles.  In addition the good $\chi^2$/DOF demonstrates that the rate
observed in each of the channels is stable.  The statitstical
uncertainty in the fit to a flat ratio of rates demonstrates an
overall rate stability of 0.01\% over the whole measurement period,
and the statistical uncertainty on each cycle's ratio of rates implies
an uncertainty per cycle of 0.06\%.  The overall rate stability easily
meets our goal of 0.03\%, however the statistical uncertainty per
cycle is not sufficient to evaluate whether short time scale
variations are present at this level.

\begin{table*}[ht]
  \caption{Area relative to the central channel $R_a$, rate relative
    to the central channel $R_r$ as fit to 114 UCN cycles,
    $\chi^2$/DOF from the fit to a constant relative rate, and
    difference between the relative area and relative rate.} \centering
  \begin{tabular}{c | c c c | c}
    \hline \hline
 Channel  & Rel. area ($R_a$)  & Rel. rate ($R_r$)  &  $\chi^2$/DOF &  $( R_a - R_r )$ \\    
    \hline 
    0  &      0.7490 $\pm$   0.0018   &    0.7065 $\pm$   0.0001    & 104.2/113 &    { }0.0425 $\pm$   0.0018\\
    1  &      0.2876 $\pm$   0.0007   &    0.2849 $\pm$   0.0001    & 119.7/113 &    { }0.0027 $\pm$   0.0007\\
    2  &      0.7634 $\pm$   0.0018   &    0.7115 $\pm$   0.0001    & 120.9/113 &    { }0.0519 $\pm$   0.0018\\
    3  &      0.3033 $\pm$   0.0006   &    0.2809 $\pm$   0.0001    & 119.4/113 &    { }0.0224 $\pm$   0.0006\\
    4  &      0.7495 $\pm$   0.0017   &    0.7481 $\pm$   0.0001    & 103.1/113 &    { }0.0014 $\pm$   0.0017\\
    5  &      0.2650 $\pm$   0.0006   &    0.2771 $\pm$   0.0001    & 132.9/113 &   -0.0121 $\pm$   0.0006\\
    6  &      0.7307 $\pm$   0.0018   &    0.7075 $\pm$   0.0001    & 100.3/113 &    { }0.0232 $\pm$   0.0018\\
    7  &      0.2672 $\pm$   0.0007   &    0.2579 $\pm$   0.0001    & 90.39/113 &    { }0.0093 $\pm$   0.0007\\
    \hline
  \end{tabular}
\label{tab:relrates}
\end{table*}

\section{ Scintillation and light-guide background simulation }\label{sec:sim}

\subsection{ Digitizer simulation }\label{sec:digisim}

In order to build probability distribution functions for scintillation
due to neutron capture, $\gamma$-ray interactions in the the
lightguides, late-light events, multiple signal events, and
combinations of signal and background events, a detailed simulation of
the voltage pulses from the lithium glass and of the digitizer PSD was
developed.  A single photo-electron (p.e.) in the PMT was assumed to
produce a Gaussian pulse with a width, $\sigma_{pe} = 6.4$~ns, an
amplitude drawn from a Gaussian with mean and width, $A=20$~mV, with a
minimum amplitude for a single p.e. of 4~mV.  The pulse width
$\sigma_{pe}$ was chosen to match the rise time of the scintillation
signal in the lithium glass.

A single scintillation signal event's pulse was then built assuming
that the arrival times of each photo-electron from the scintillation
signal followed a rise time, $\tau_R=6.4$~ns, a fast scintillation
fall time, $\tau_F=41.7$~ns, and a slow scintillation fall time,
$\tau_S=2000$~ns.  The probability, $P(t)$, of having a photo-electron
at a given time, $t$, when the scintillation light starts arriving at
time, $T$, was drawn from the Probability Distribution
Function (PDF):
\begin{equation}
  P(t) = \left\{
    \begin{array}{l}
      A ( 1 - {\tt e}^{-(t-T)/\tau_R } ),\ \ T<t<T+5\tau_R \\
      A ( (1-f_L) {\tt e}^{ -(t-T-5\tau_R)/\tau_F } +   \\
      \ \ f_L {\tt e}^{ -(t-T-5\tau_R)/\tau_S } ),\ \  t>=T+5\tau_R.
    \end{array}
    \right.
\end{equation}

The number of photo-electrons for a single neutron event was drawn
from a Poisson distribution with a mean number of photo-electrons of
83.  The fraction of the scintillation light in the late-light was
$f_L=1$\%\cite{tmurata2012}.  

\begin{figure}[!htpb]
\centering
\includegraphics[width=0.45\textwidth]{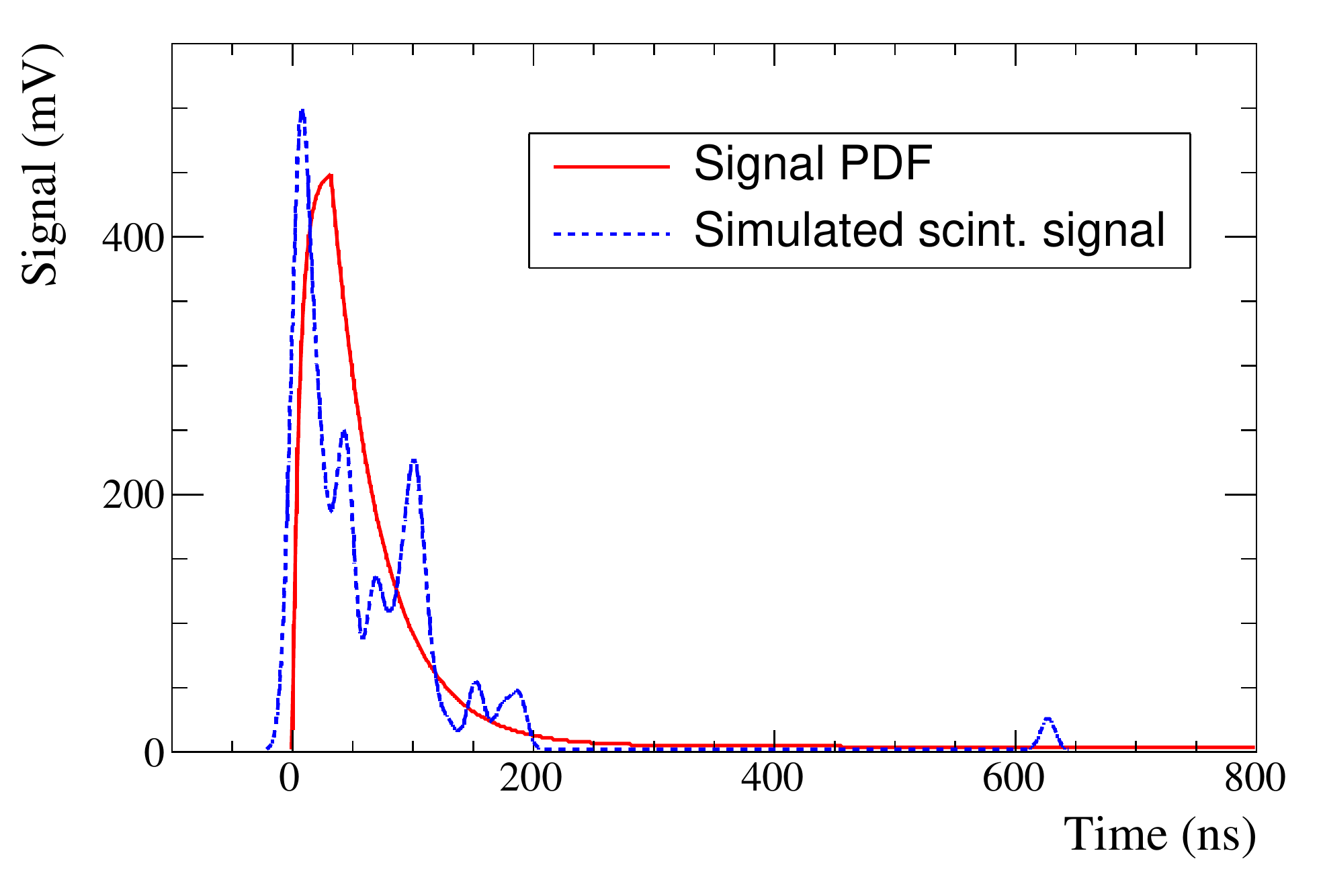}
\includegraphics[width=0.45\textwidth]{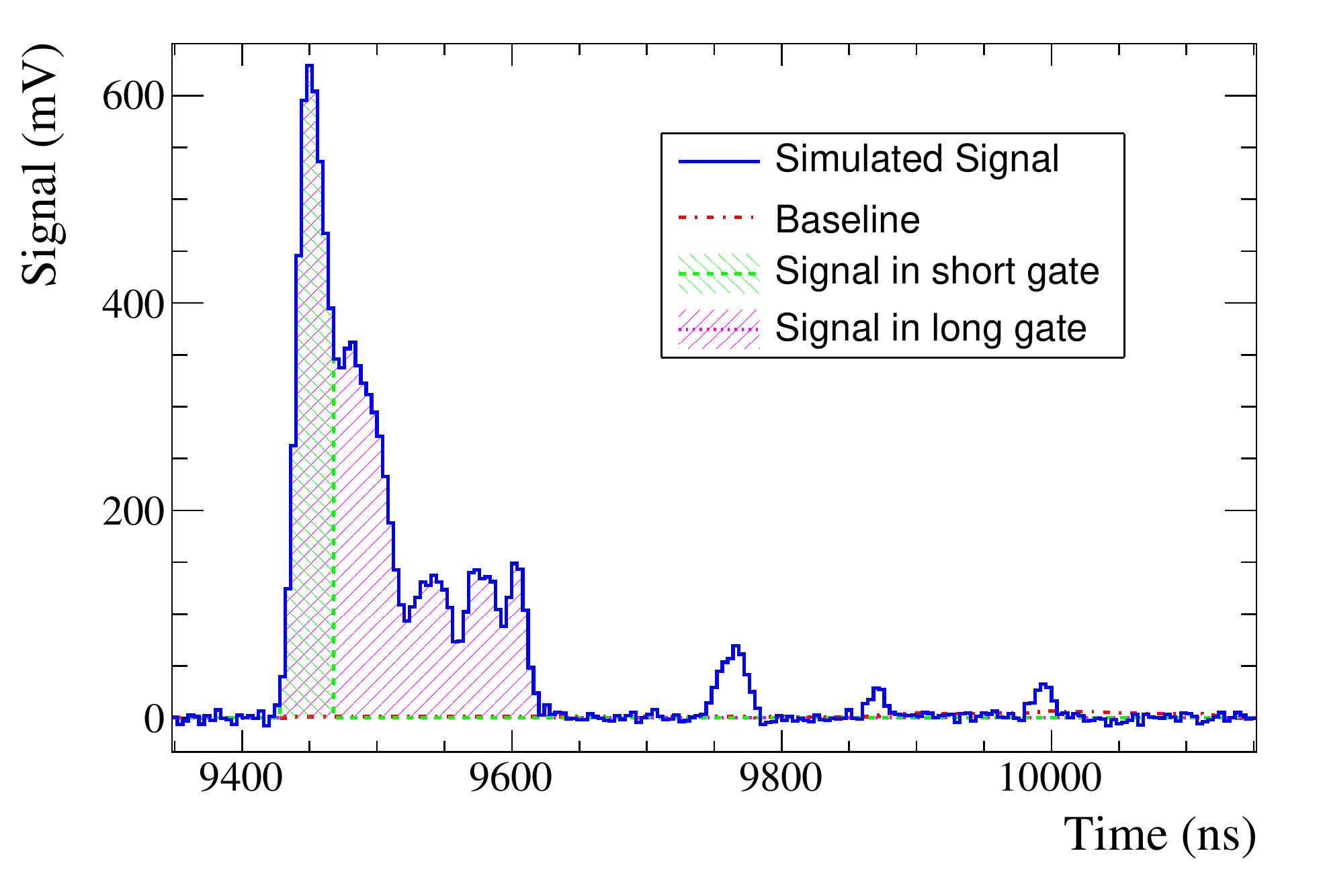}
\caption{ Simulated scintillation signal (blue dashed) and
  distribution function used to generate it (red solid) is shown in
  the top panel.  The bottom panel shows the digitizer treatment of
  another simulated signal (solid line), where the baseline
  calculation (red dot-dashed line), signal within the short gate
  (green right-diagonal fill) and signal within the long gate (magenta
  left-diagonal fill) are shown (colour online).}\label{fig:signalpdf}
\end{figure}

All of the values used in the simulation, as described above, were
chosen to best match the PSD and $Q_L$ distributions in the data.
This tuning was done by plotting the mean, sigma and mean over sigma
of the $Q_L$ and PSD distributions in single dimension scans of the
simulation parameters until the distributions matched these same values
from the data distributions.  The distribution of scintillation
photo-electron arrival times along with a sample pulse is shown in the
top panel of Fig.~\ref{fig:signalpdf}.  Note that positive pulses with
a threshold above a baseline were used in our simulation, while in the
data the pulses are negative, and the threshold is some number of ADC
below the baseline.

The number of photo-electrons for the background signal was drawn from
an exponential distribution with an average of $7.5$ photo-electrons.
This distribution was chosen to match the dominant component of the
background observed in data.

\begin{figure}[!htpb]
\centering
\includegraphics[width=0.45\textwidth]{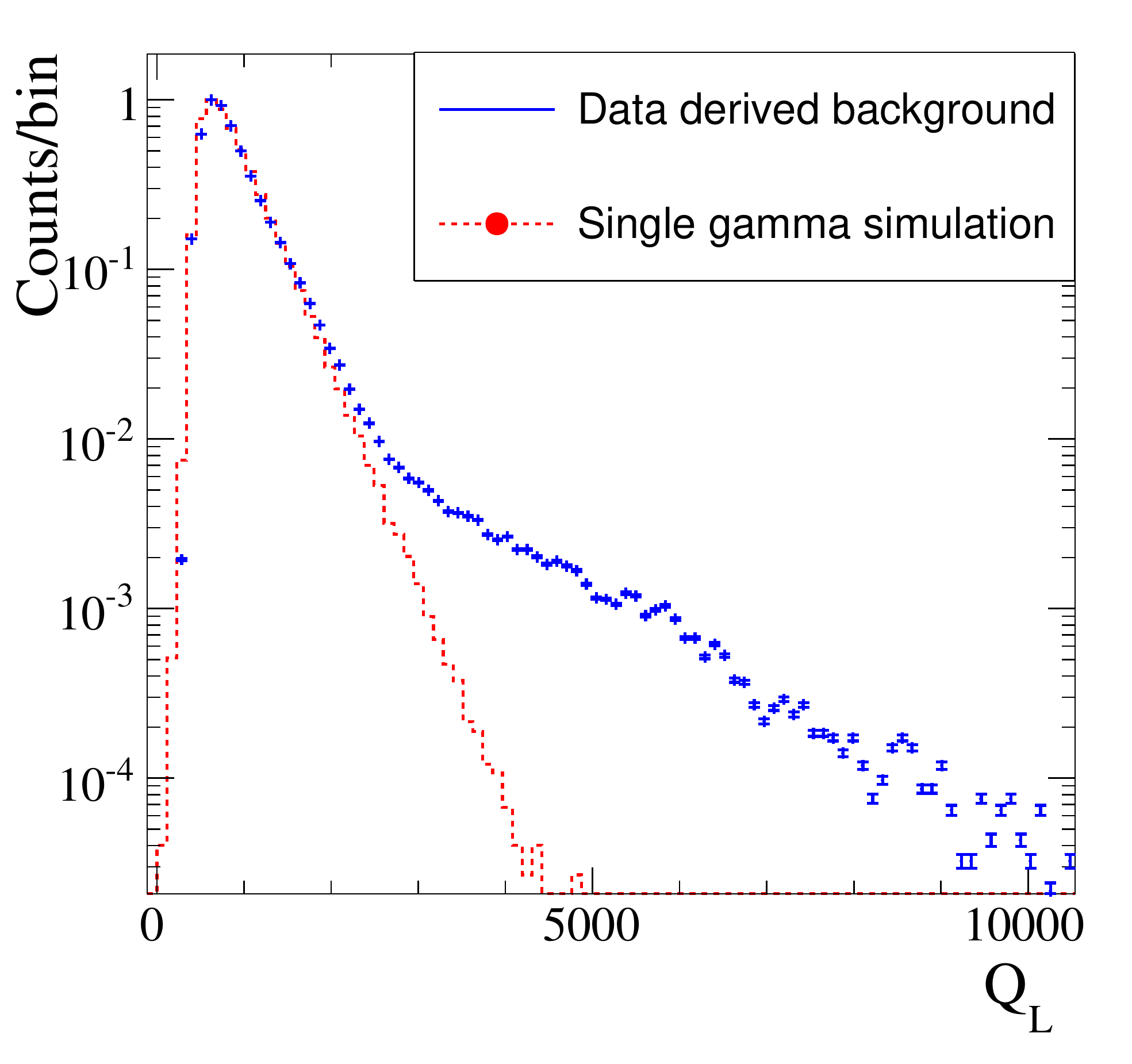}
\caption{ Background counts as a function of $Q_L$ from data taken
  during periods without UCN are shown as blue crosses, and is
  overlaid with the single gamma ray background shown as a red-dashed
  line (colour online).}\label{fig:gammamatching}
\end{figure}

The matching of the single gamma background from the simulation and
from data during times without UCN is shown in
Fig.~\ref{fig:gammamatching}.  The data distribution is used to
represent the single gamma ray background from interactions in the
scintillator and light-guide, as well as backgrounds from thermal
neutrons.  The distribution in the data has more counts at high $Q_L$
due to the presence of scintillation events in the lithium glass due
to gamma-ray and thermal neutron interactions.  The gamma ray
background simulation is used for the simulation of pile-up of the
gamma signals with themselves, and with the neutron signals.

These simulated data were then sent to a digitizer simulation
described in the following section.  The simulation of the pulses was
used to generate 0.1 second long sets of data where the signal and
background pulses were generated at a specified random rate.

\subsection{ Probability distribution functions from the simulations } 

Seven categories of events are considered.  Self explanatory
categories are single neutrons (1n), single backgrounds (1$\gamma$),
pile-up of multiple signal neutrons (Nn), pile-up of multiple
backgrounds (N$\gamma$), and pile-up of a single neutron with a
background (1n1$\gamma$).  After the end of a triggered event the
digitizer channel is busy for 150~ns beyond the end of the long gate.
If a neutron comes during this dead-time, part of its charge is not
collected.  We call these events single neutrons during the deadtime
(1n deadtime). The last event type constitutes triggers on late-light
from the scintillator (0n0$\gamma$).

\begin{figure}[!htpb]
\centering
\includegraphics[width=0.45\textwidth]{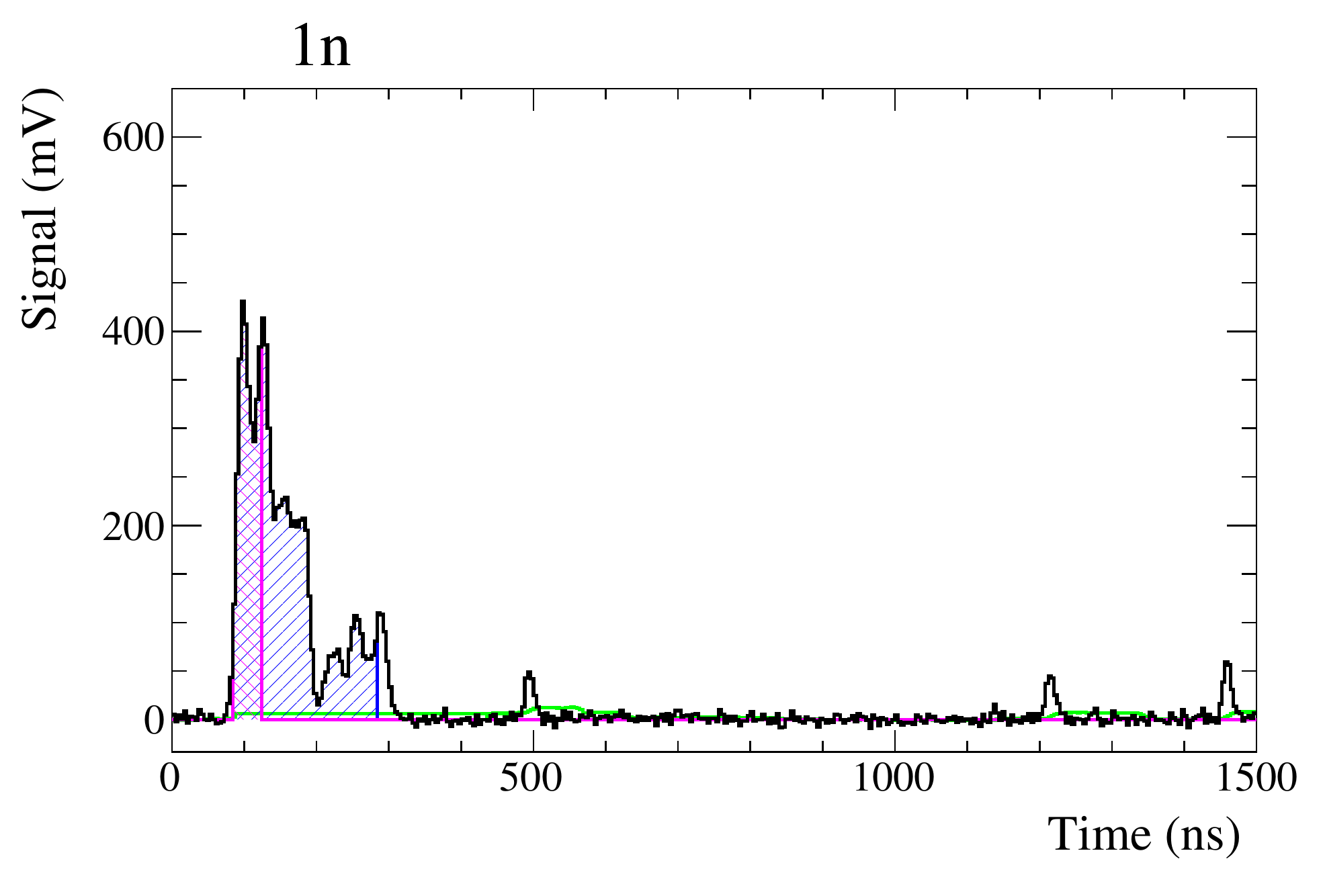}
\includegraphics[width=0.45\textwidth]{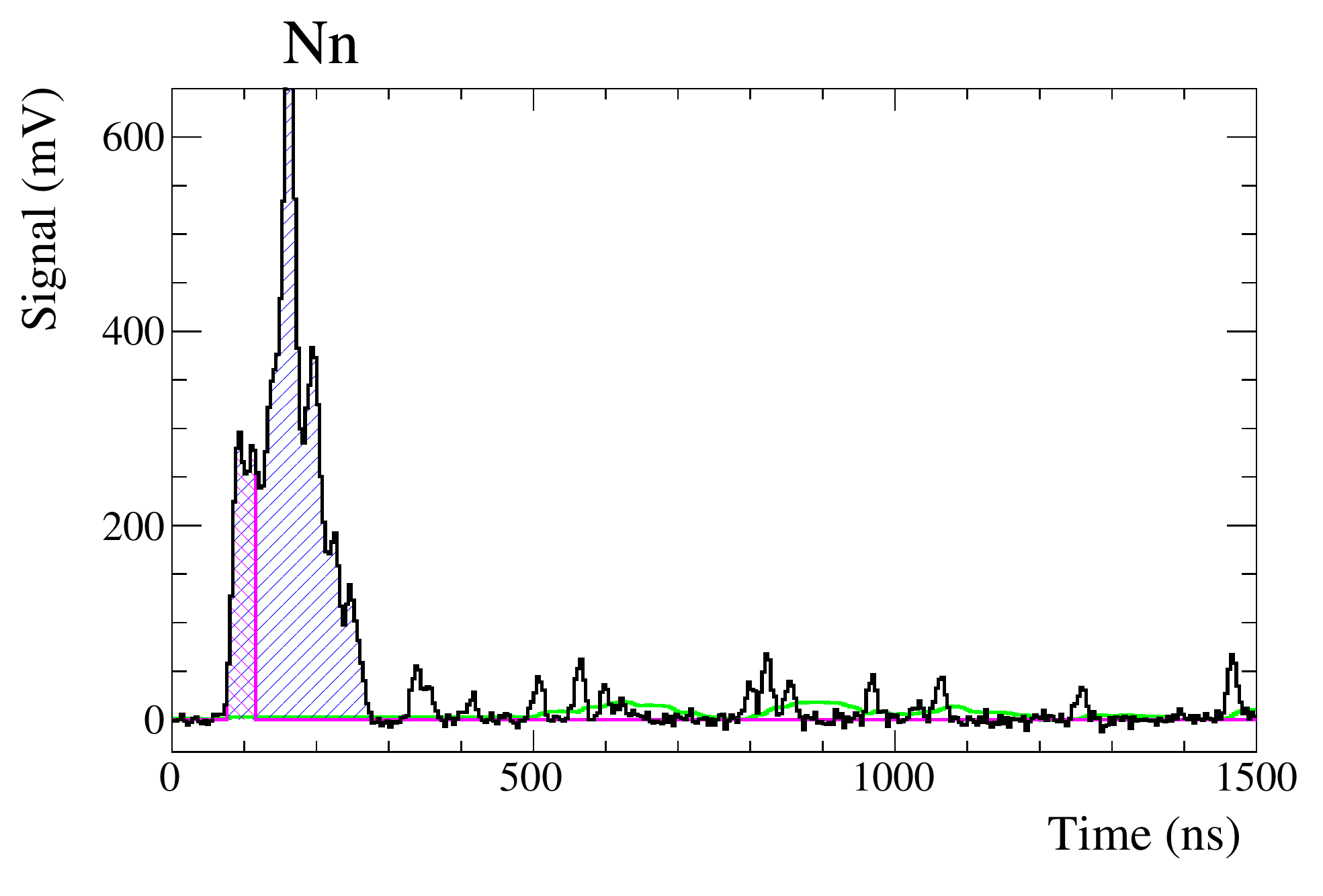}
\includegraphics[width=0.45\textwidth]{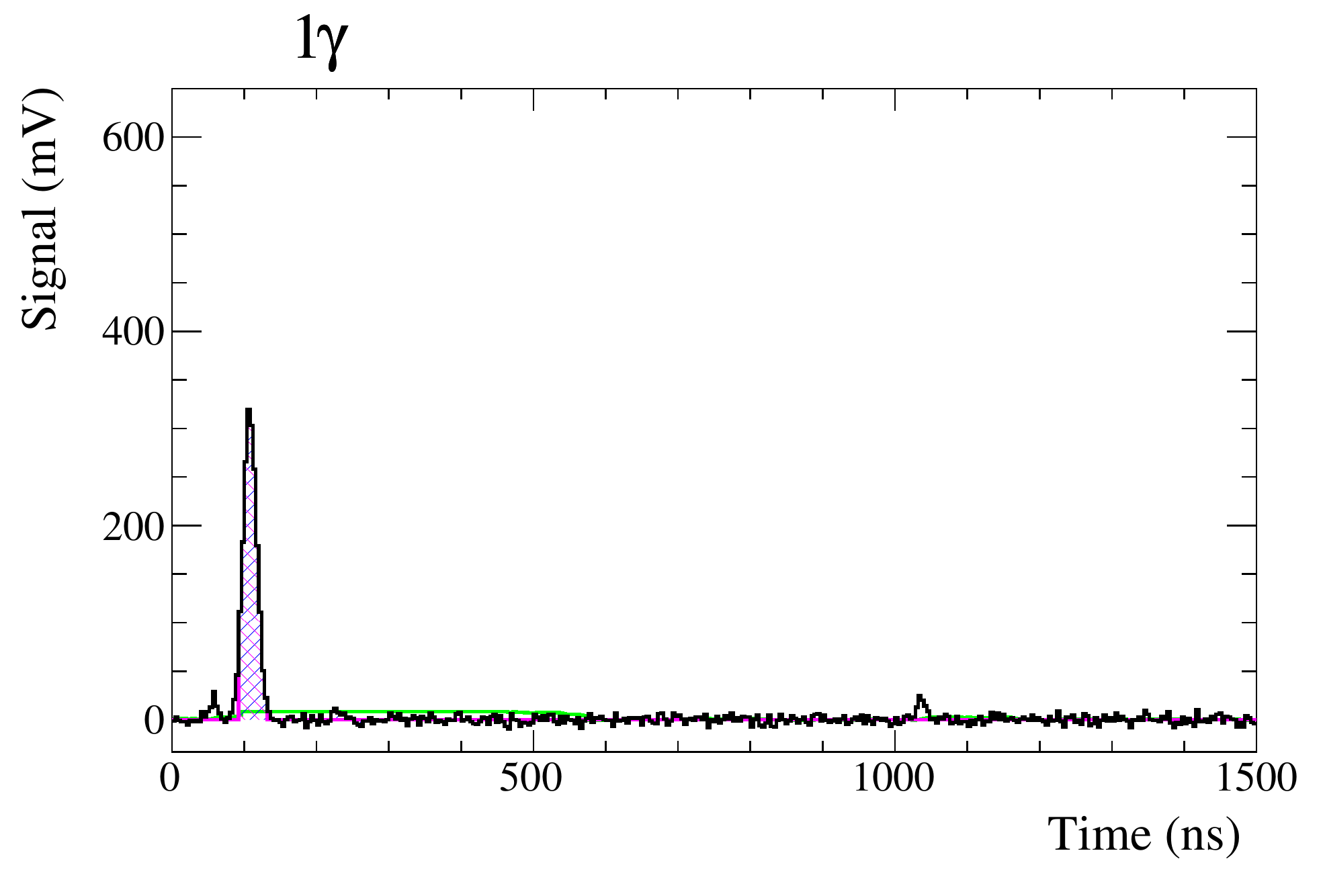}
\includegraphics[width=0.45\textwidth]{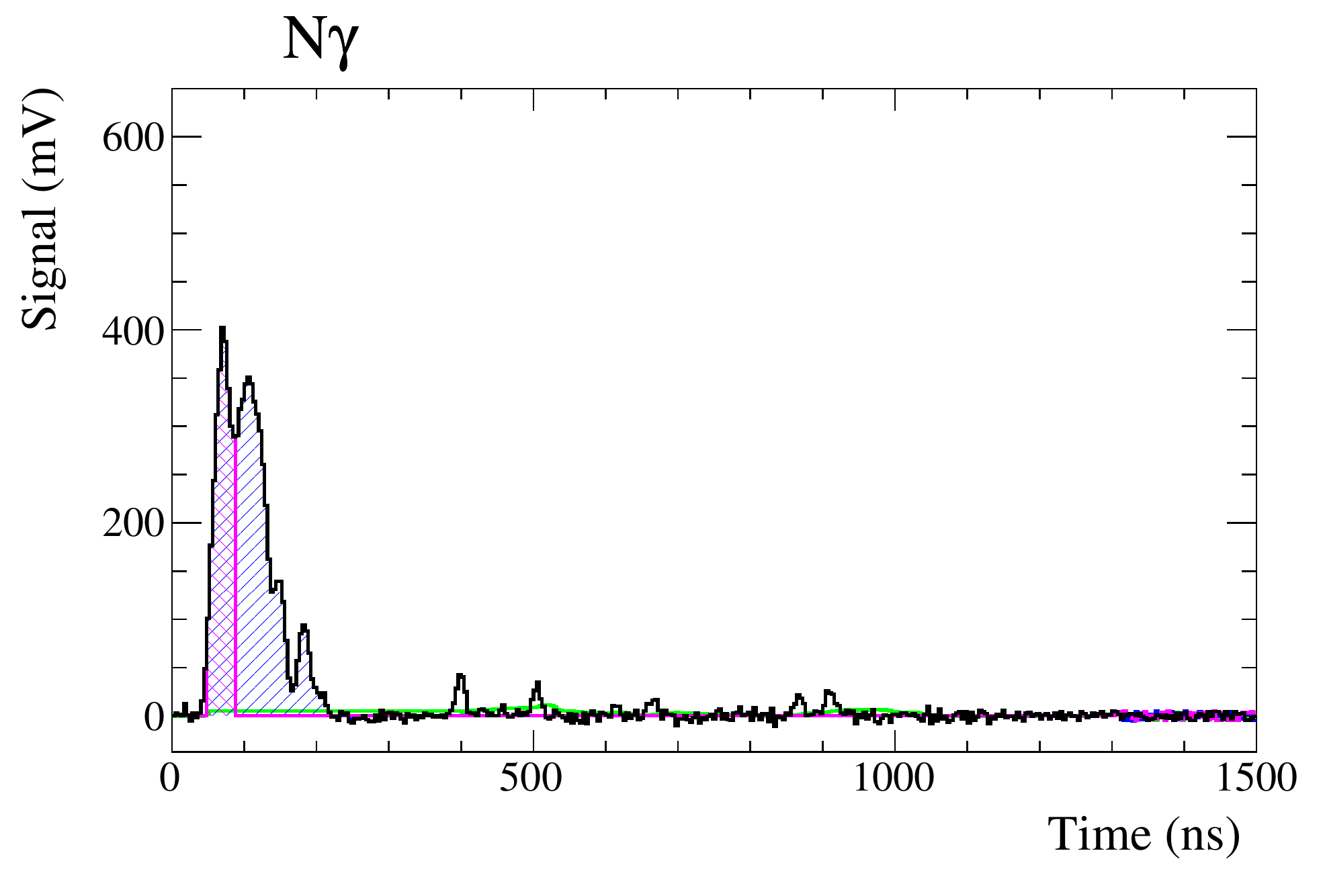}
\caption{ Simulated signals for different combinations of signal and
  background events.  The magenta left-diagonal-hatched region
  represents the $Q_S$ portion of the signal, the blue
  right-diagonal-hatched region represents $Q_L$, and the green line
  (colour online) represents the average baseline.  From top to bottom
  the plots show: 1n, Nn (2 neutrons), 1$\gamma$, and N$\gamma$ (3
  $\gamma$) (colour online). }\label{fig:eventTypes}
\end{figure}

\begin{figure}[!htpb]
\centering
\includegraphics[width=0.45\textwidth]{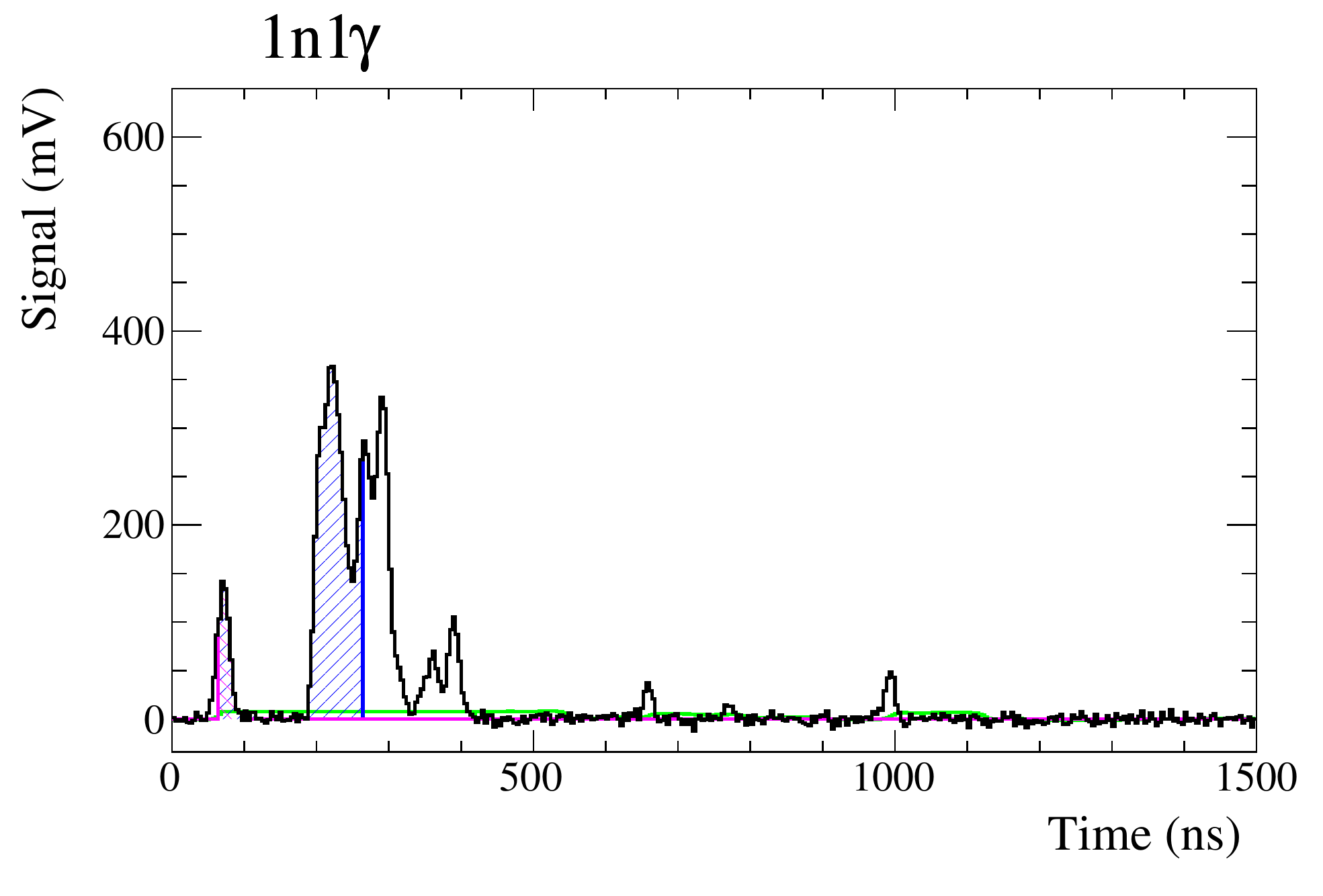}
\includegraphics[width=0.45\textwidth]{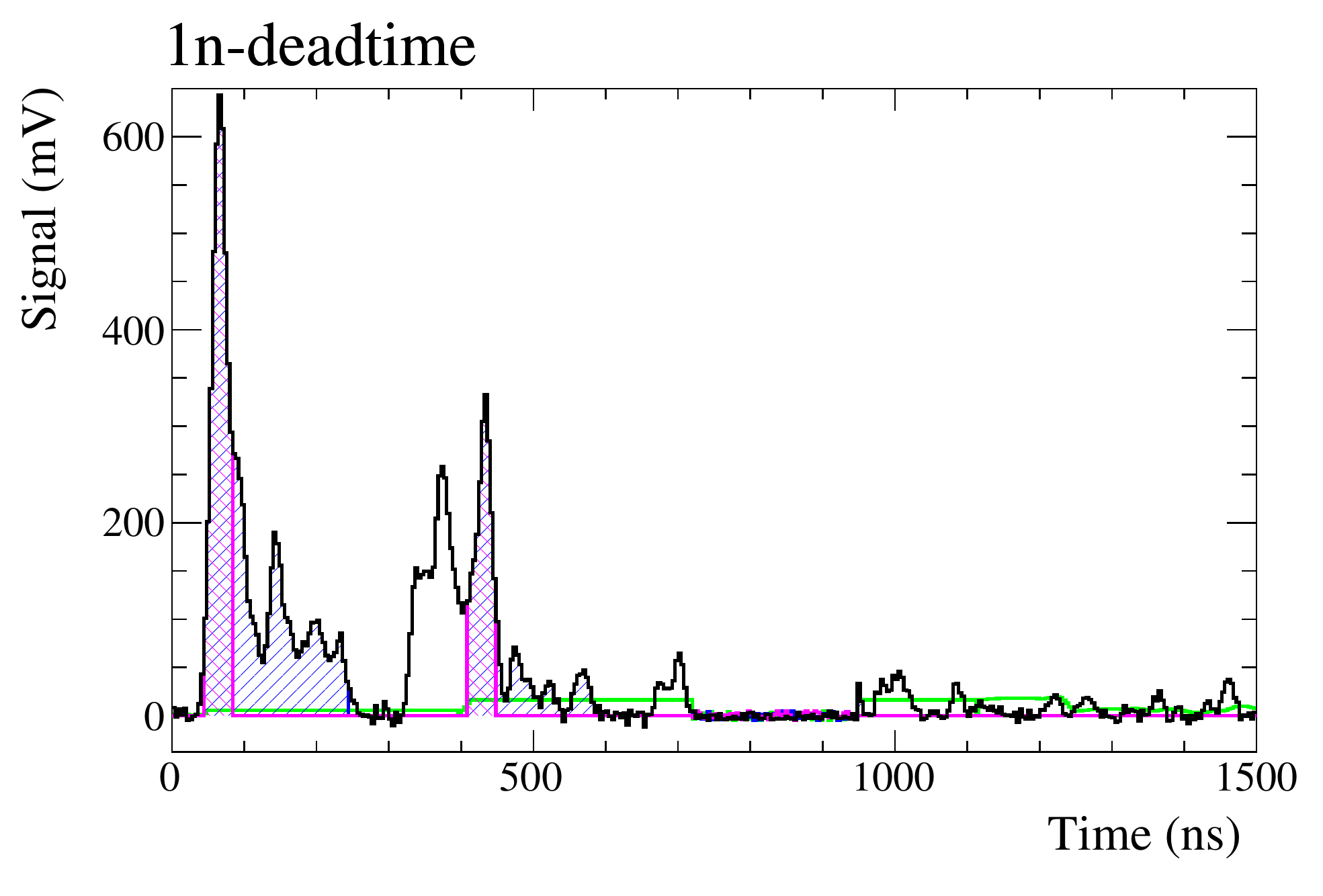}
\includegraphics[width=0.45\textwidth]{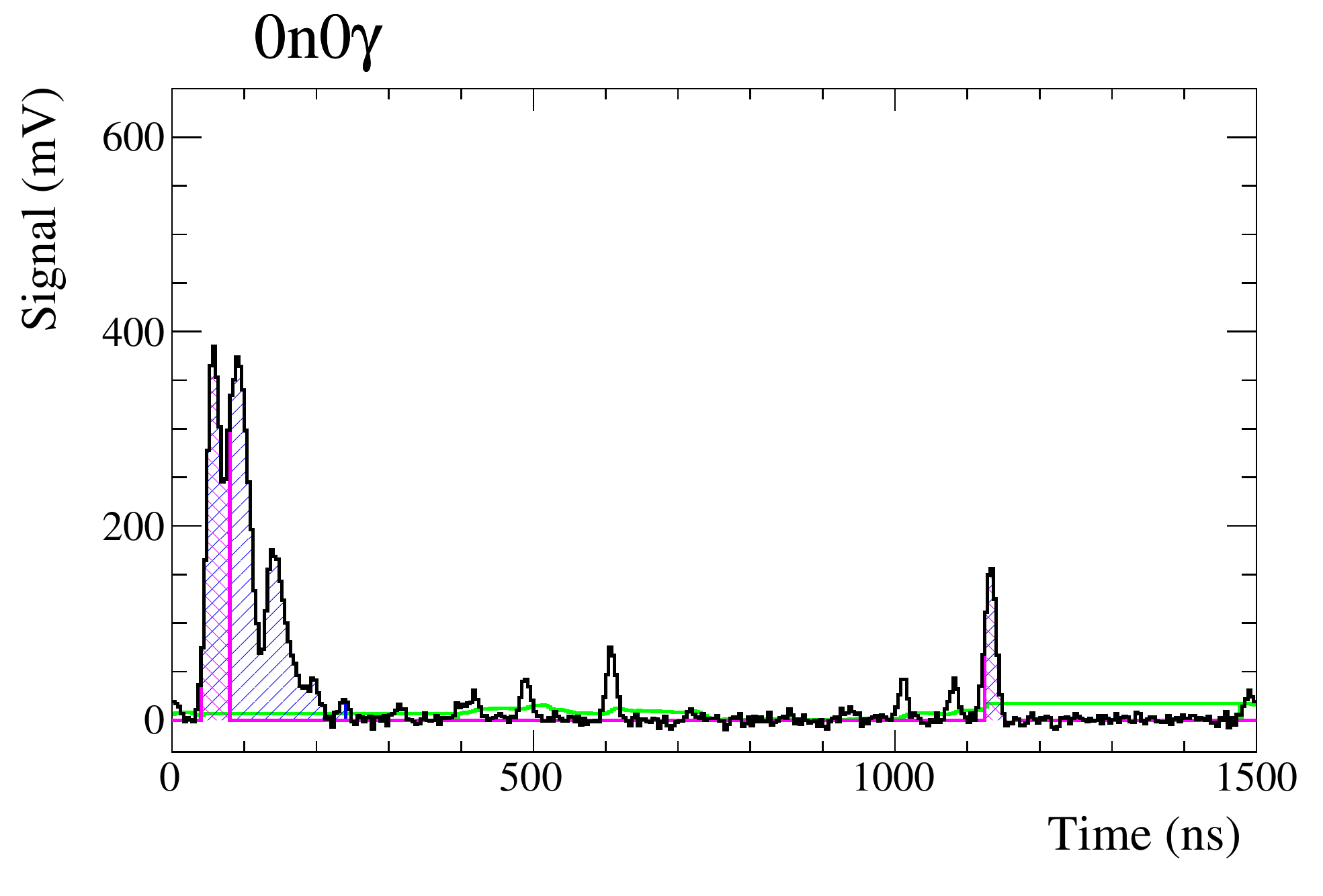}
\caption{ From top to bottom these plots show: a single signal plus
  single background event, a dead-time event, and a late-light
  re-trigger event (colour online). }\label{fig:eventTypes2}
\end{figure}

Simulated electronic pulses for each of these possible combinations
are shown in Fig.~\ref{fig:eventTypes} and Fig.~\ref{fig:eventTypes2}.

The combination of the signal and background pulse simulations with
the digitizer simulation is used to generate PDFs in the PSD versus
$Q_L$ space for single neutron events, gamma events and different
possible combinations pile up of events.

In the simulation, a a fixed (random) rate of 10~kHz of $\gamma$
interactions in the light-guides and 10~kHz of neutron interactions in
the scintillator has been chosen.  We separate out triggers that are
from single interactions and put them in singles PDFs (1n, or
1$\gamma$ ).  Triggers that have multiple interactions in the
long-gate are put in multiples PDFs (Nn, N$\gamma$, and 1n1$\gamma$).
By allowing the normalizations of these PDFs to be changed we
approximately account for the varying rate during the UCN cycle.

In the case of the single background event, it was possible to use the
data from cycles when no UCN are produced, as shown in
Fig.~\ref{fig:gammamatching}.  In our model we are assuming that
background components from thermal neutrons and gamma-rays interacting
in the scintillators are present in this PDF derived from data.  The
rest of the PDFs are derived from the simulations described above.

\begin{figure}[!htpb]
\centering
\includegraphics[width=0.45\textwidth]{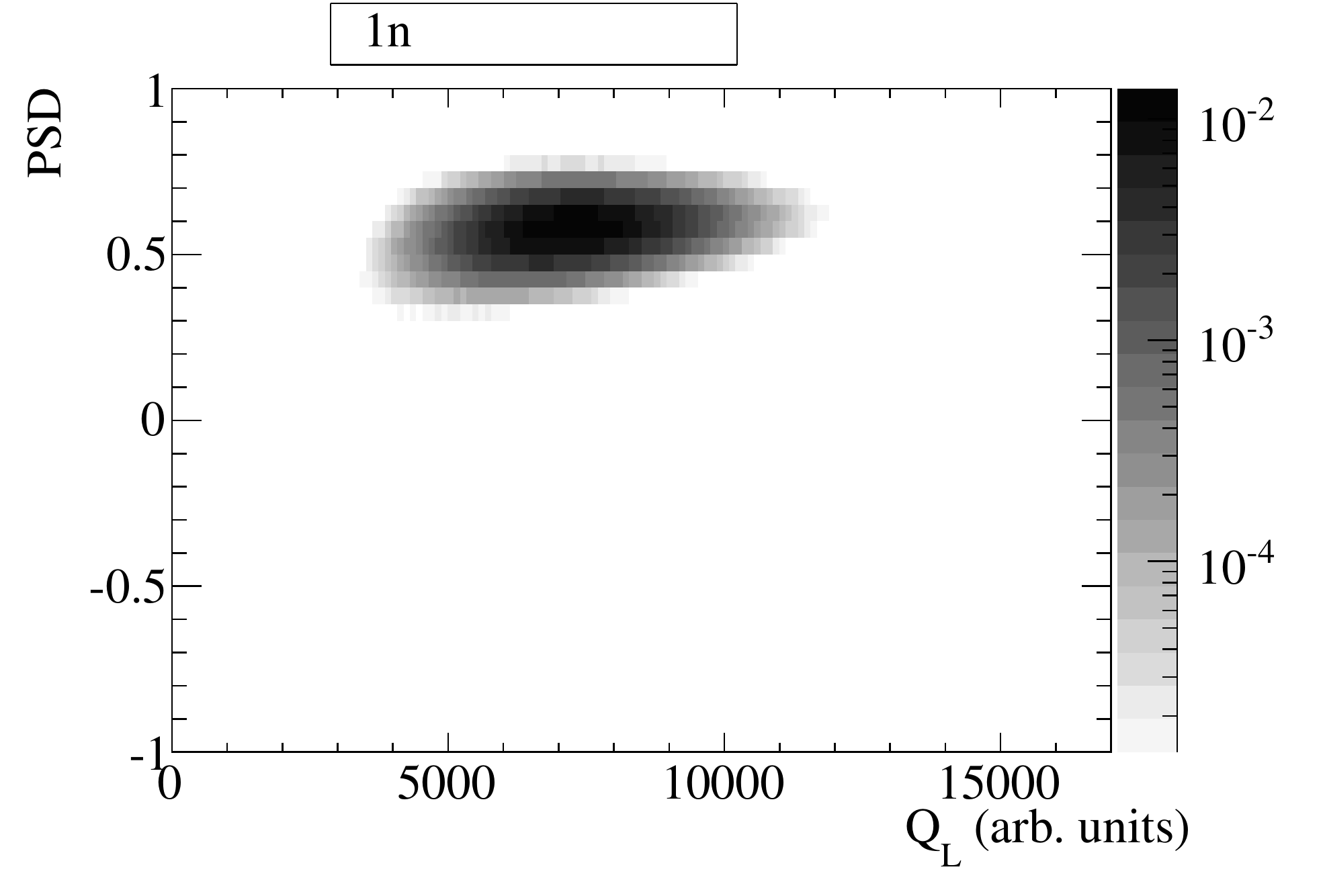}
\includegraphics[width=0.45\textwidth]{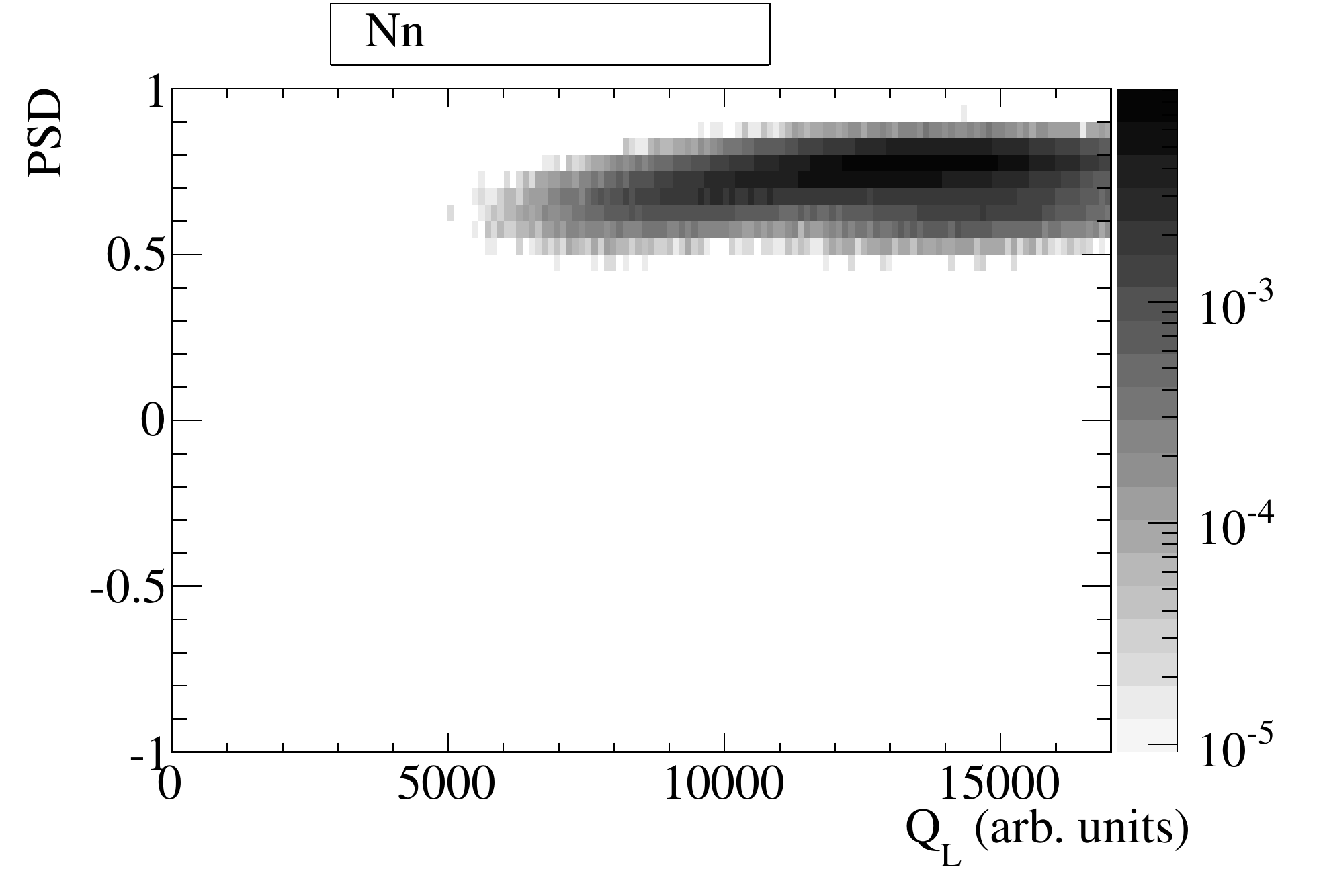}
\includegraphics[width=0.45\textwidth]{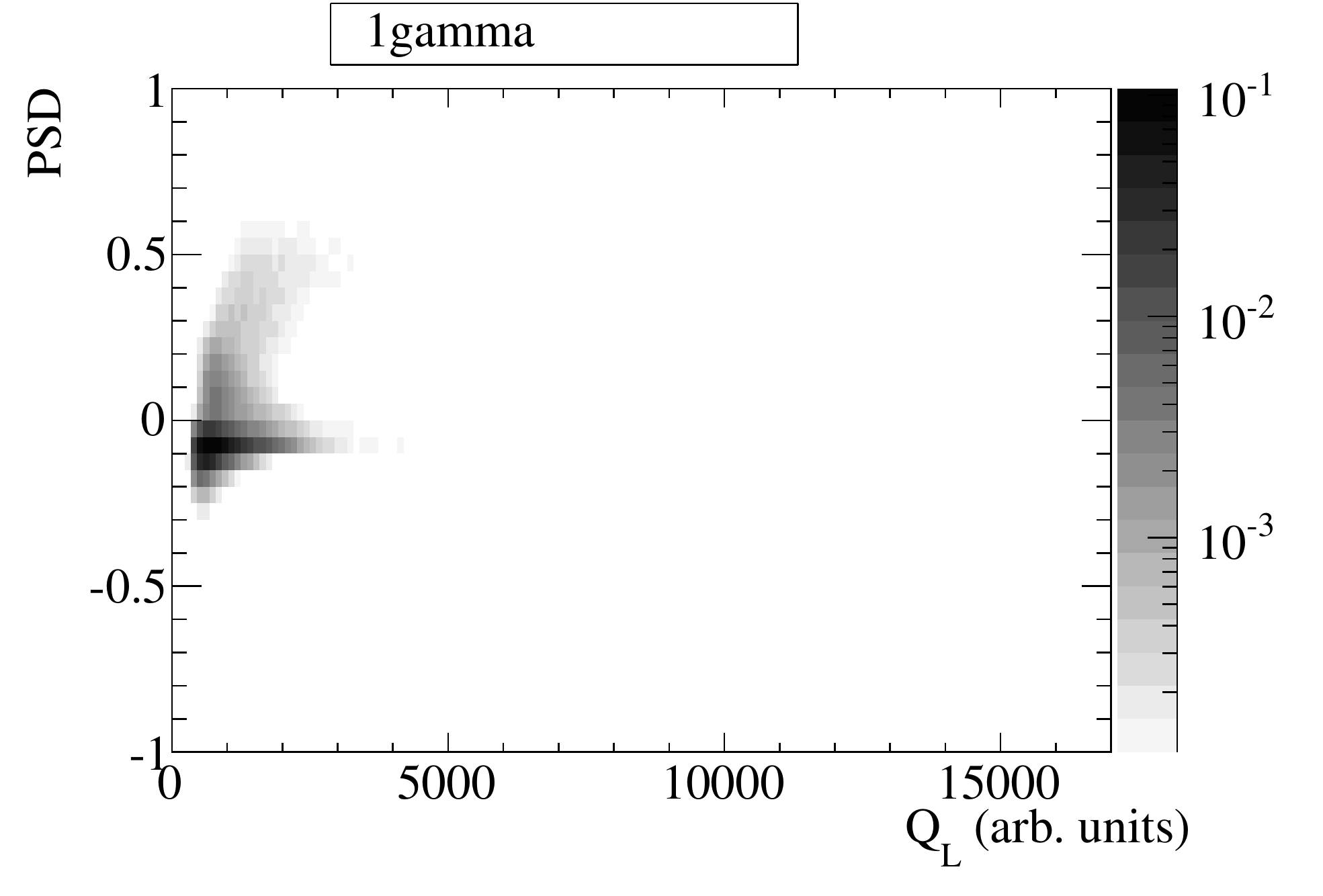}
\caption{ Event count (or relative event count) as a function of PSD
  and $Q_L$. From top to bottom: single neutron simulation, multiple
  neutron simulation, and background (from data from cycles without
  proton beam).} \label{fig:eventSpectra}
\end{figure}

\begin{figure}[!htpb]
\centering
\includegraphics[width=0.45\textwidth]{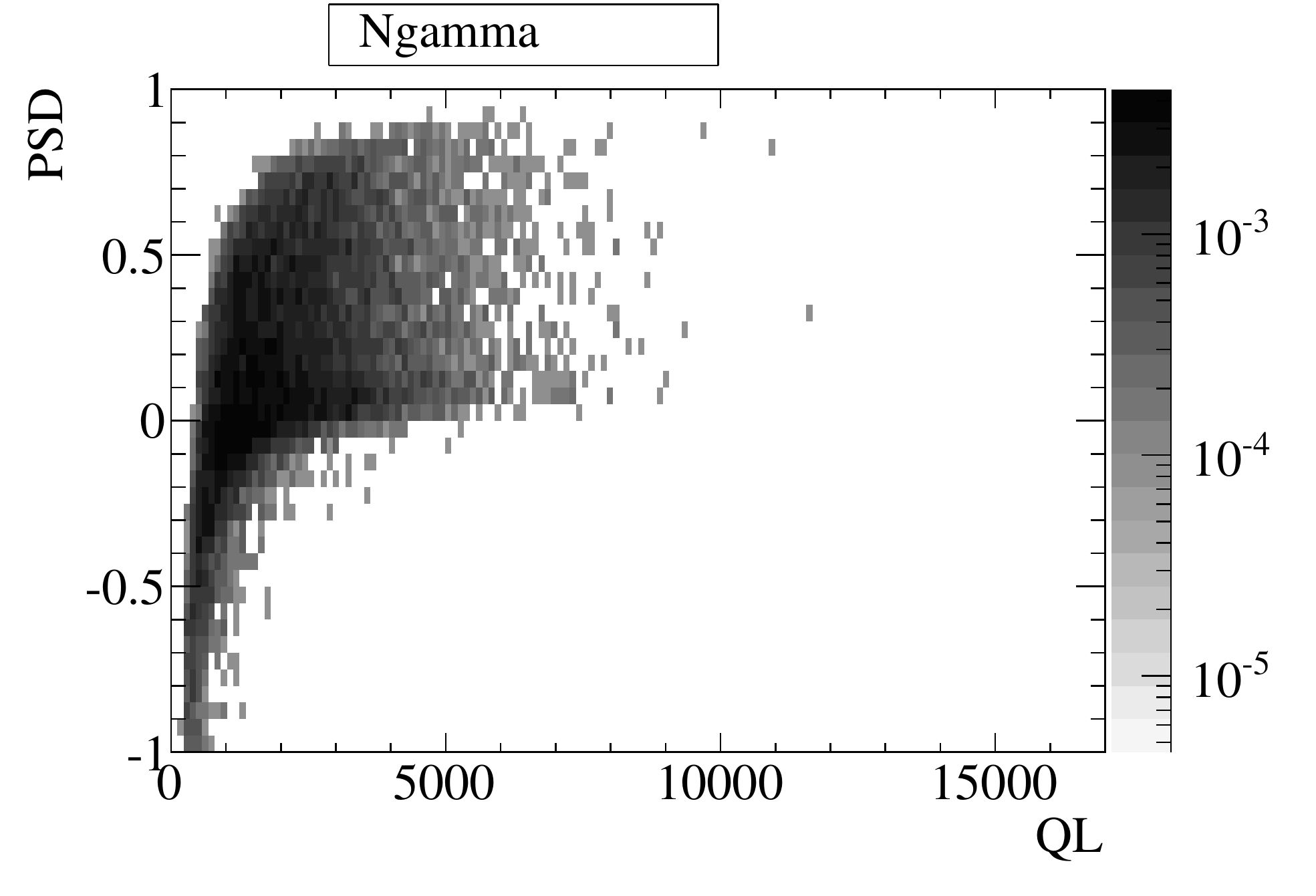}
\includegraphics[width=0.45\textwidth]{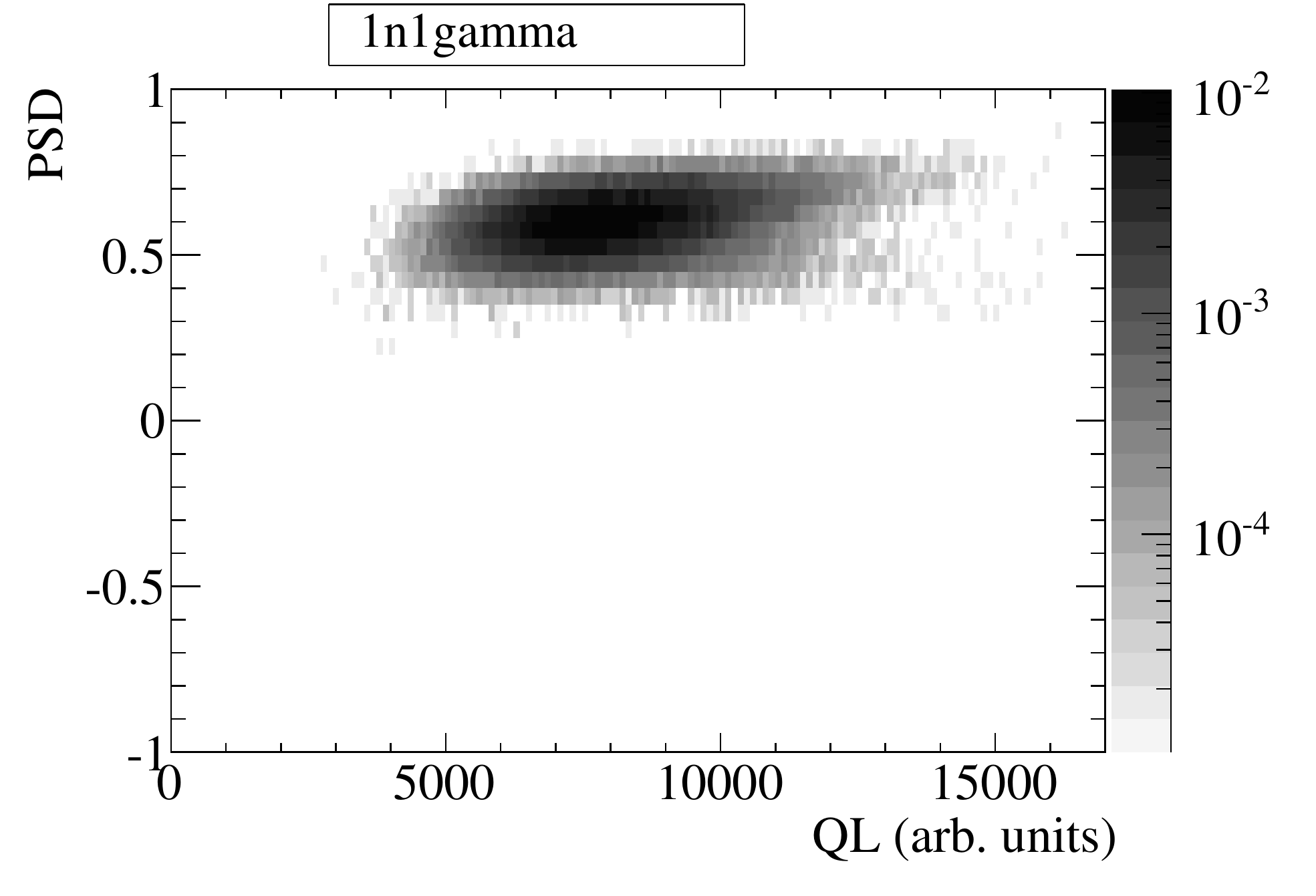}
\includegraphics[width=0.45\textwidth]{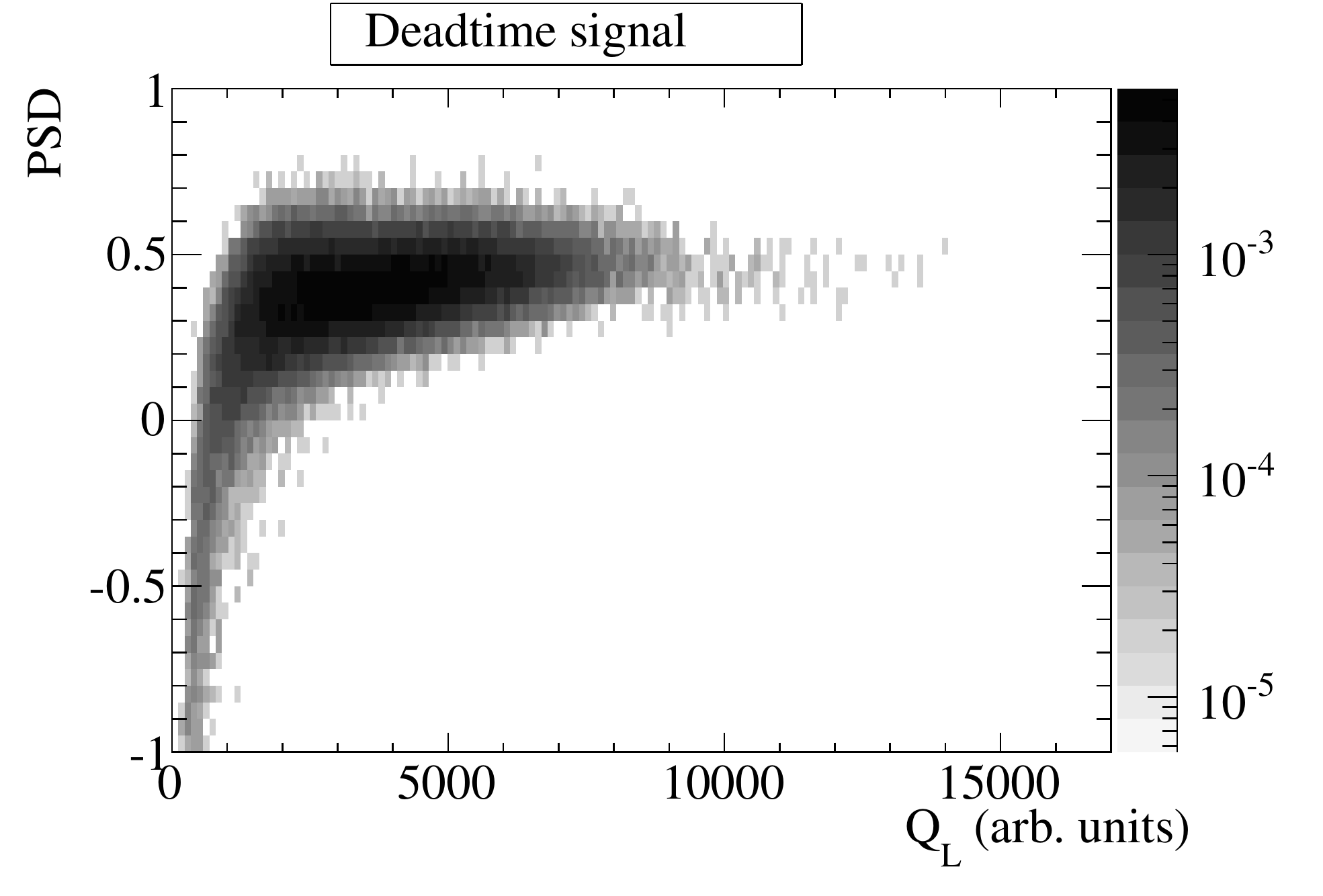}
\includegraphics[width=0.45\textwidth]{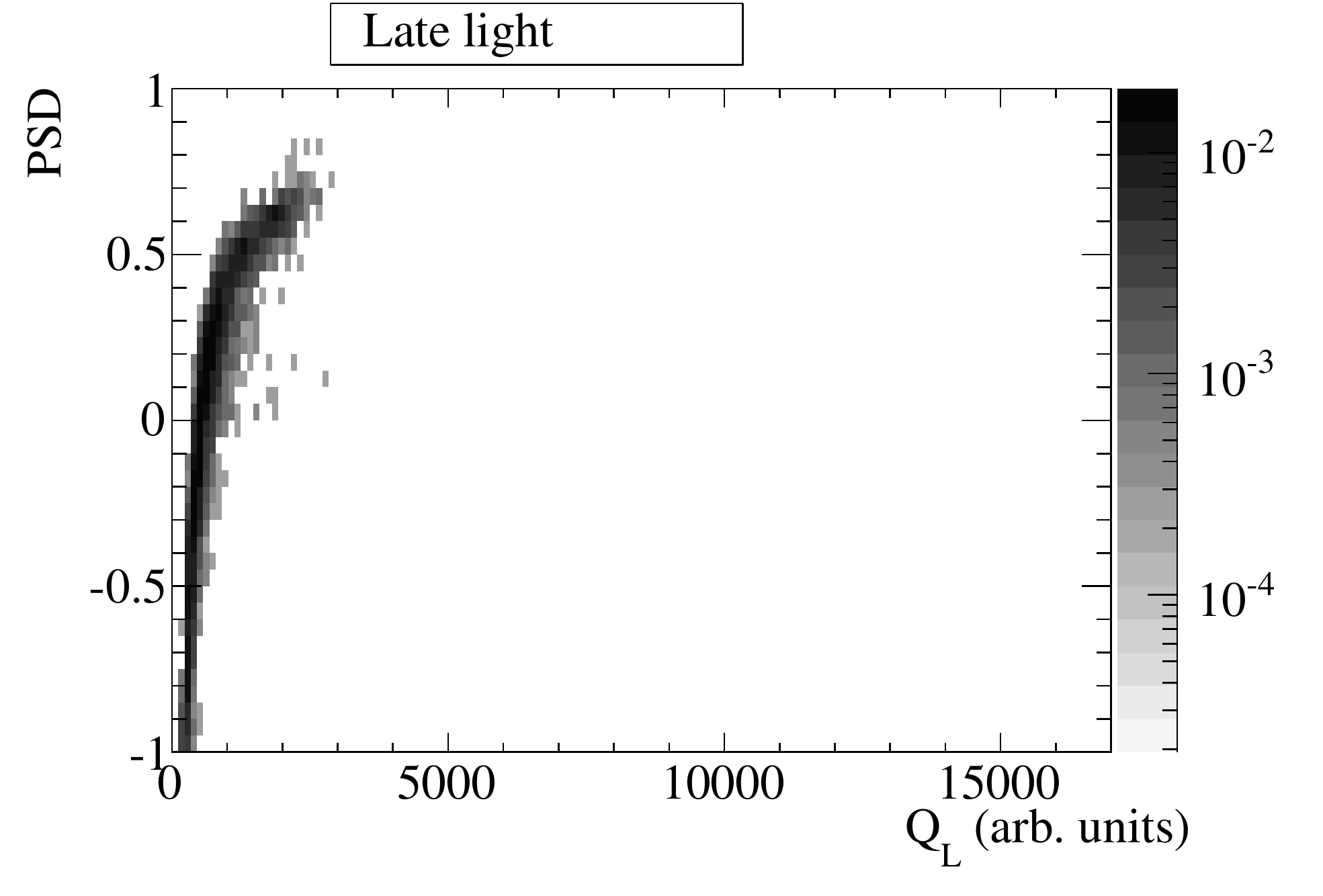}
\caption{ Event count (or relative event count) as a function of PSD
  and $Q_L$.  From top to bottom: multiple background simulation,
  single neutron plus single background simulation, dead-time neutron
  simulation, and re-trigger on late-light simulation.}
\label{fig:eventSpectra2}
\end{figure}

The PDFs for the different combinations of signal and background as
found by the simulation are shown in Fig.~\ref{fig:eventSpectra} and
\ref{fig:eventSpectra2}.  

The single neutron (1n) pulses extend out to the long gate integration
time, meaning that their $Q_L$ will be larger than $Q_S$.  The PSD is
therefore greater than zero, and in this case is observed to be
$\sim$0.5.  A pile-up of two or more neutrons (Nn) has pulses at least
as large as a single neutron, and can extend out to later time
depending on the time separation between the two neutrons within the
long integration time.  The Nn events therefore have larger $Q_L$ than
the 1n events, while $Q_S$ is unchanged.  The Nn events therefore, on
average, have a slightly higher PSD than the 1n events.

The single gamma (1$\gamma$) pulses fit within the short integration
time, therefore they have similar $Q_S$ and $Q_L$ values.  The PSD for
these events is near zero.  In fact, due to the settings of the
digitizer, the PSD is slightly below zero due to the baseline shifting
slightly down before the trigger time.  This negative PSD effect
becomes even more pronounced when looking at N$\gamma$, 1n deadtime,
and 0n0$\gamma$ events where the baseline is shifted low due to the
previous pulse on the channel.  The 1n1$\gamma$ sample looks fairly
similar to the 1n distribution, but there is some tail to higher
$Q_L$ due to the additional light from the $\gamma$.

The 1n deadtime events do not integrate the full charge from the
neutron pulse due to the trigger coming too late.  For that reason the
$Q_L$ for these events is lower than for the 1n events.  These
triggers come soon after the previous trigger, so the baseline is
sometimes shifted low due to possible late-light of the previous
pulse.  The baseline shift can cause $Q_L$ to be lower than $Q_S$
causing some of these events to have negative PSD values.

Finally, the late-light pulses have low $Q_L$ and low $Q_S$, since
they are triggers just above threshold.  The small charges, and
possible baseline shift leads to these events having a wide range of
PSD values.

The data contain all of these different event categories, and some
combination of these PDFs fills in the PSD versus $Q_L$ distribution
observed in data. A template fit of these PDFs to estimate amount of
signal and background in the data is possible.

\section{Estimate of the absolute detector efficiency and background contamination}\label{sec:eff}  

To determine the $^{6}$Li detector's UCN detection efficiency we
consider the active area of the detector, losses due to absorption in
the front face including the $^{7}$Li(n,$\gamma$)$^8$Li reaction, and
the transmission of the $^{6}$Li depleted layer.  Neutrons lost due to
the background rejection cuts are also considered.  The effects of
surface impurities is not straight forward to estimate and may affect
the efficiency.  The effects of surface impurities is estimated to be
bounded by the 5\% differences in relative rates of the channels
presented in Section~\ref{sec:chantochan}.  The spectrum of UCN that
are being detected also needs to be taken into account.  Here we will
assume that we want to know the efficiency for detecting UCN with
kinetic energy far enough above the effective Fermi potential of the
GS20 lithium glass ($\sim$ $103.4$~neV) so that we can neglect UCN
reflection.  The efficiency for detecting UCN with a known energy
spectrum could then be estimated using a model for UCN reflection from
a Fermi potential of 103.4~neV.  Note that the Fermi potential of the
GS30 (83~neV) is lower, and so should have negligible effect.

\subsection{ Detector effective area }

\begin{figure}[!htpb]
\centering \includegraphics[width = 0.45\textwidth]{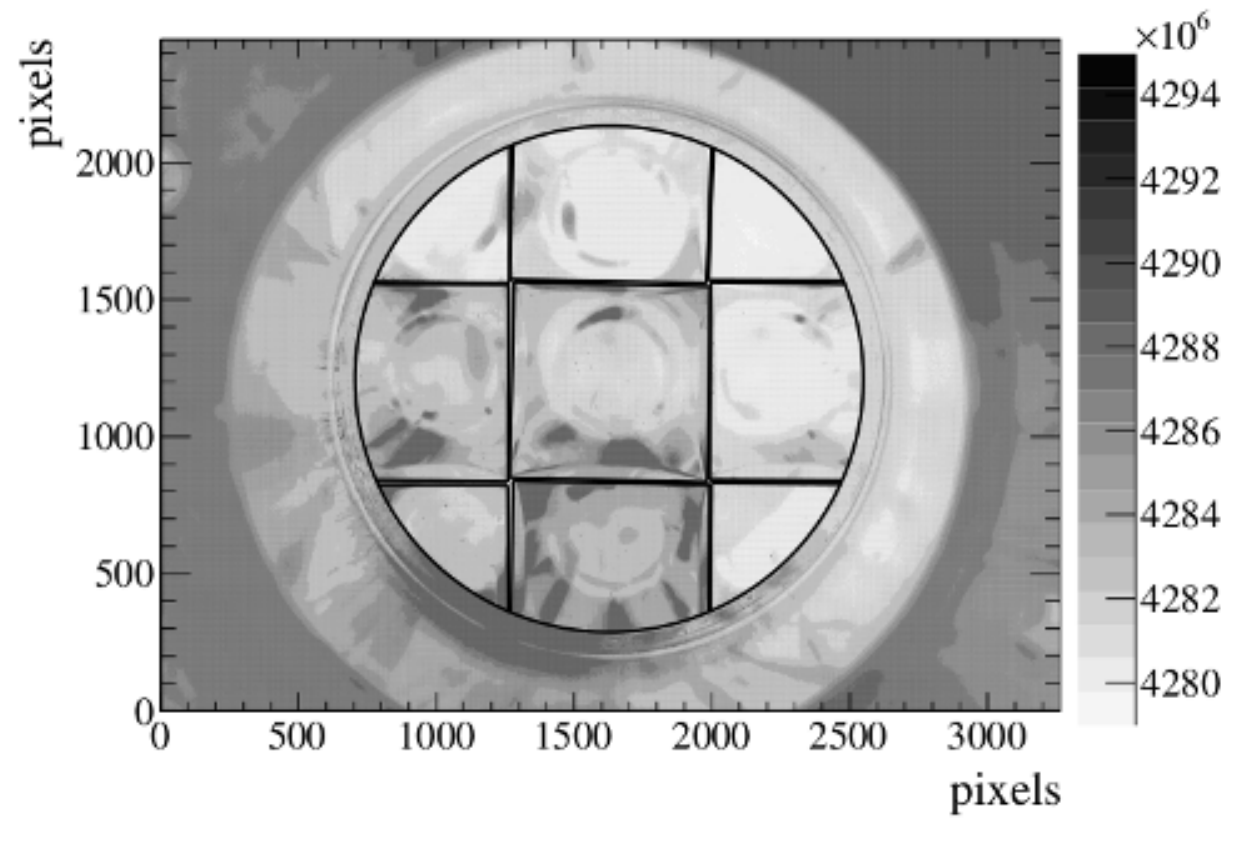}
\caption{Picture of the detector face used for estimating the
  areal efficiency of the detector.  The detector aperture and edges
  of the lithium glass are identified by the black lines.}
\label{fig:detface}
\end{figure}

The estimate of the detector's effective area comes from a photograph
of the detector's front face, where the side length of each $^{6}$Li
glass tile is $29.0\pm0.1$~mm.  Using this length the number of pixels
in the photograph that make up the circular aperture of the detector
are counted as the denominator, and the count of pixels containing
$^{6}$ Li glass tiles as the numerator.  From the photograph of the
detector face shown in Fig.~\ref{fig:detface} in gray scale, the
active area of the detector is estimated to be $97.4\pm0.1$\%.

\subsection{ Estimation of UCN absorption in Li depleted layer }

Measurements of the transmission of UCN through different thicknesses
of GS30 have been conducted by the LPC Caen group at the Institute
Laue-Langevin in Grenoble~\cite{gban2016}.  Using their measurements
for a $55\pm10$~$\mu$m GS30 layer the UCN transmission is
$92.6^{+1.2}_{-1.8}$\%. The uncertainty is asymmetric due to the
exponential nature of the attenuation through a layer with uncertain
thickness.  These results depend on the surface roughness and impurity
present in the detectors of the LPC Caen group.  Our detector's
lithium glass and surfaces were prepared by the same companies used by
the LPC Caen group and therefore forms a good estimate for UCN losses
in trasmission trough the depleted layer.

Part of the losses in the depleted layer occur due to the 6.6\% of
$^{7}$Li in the layer which allows neutron capture via
$^{7}$Li(n,$\gamma$)$^{8}$Li.  Assuming a $1/v$ law for neutron cross
sections from cold down to 100~neV UCN energy we estimate the UCN
capture cross section on $^{7}$Li to be $\sigma \sim
6.0$~barns~\cite{mheil1998}.  The effect is therefore fairly
negligible since the fraction of UCN making it through $55$~$\mu$m
GS30 due to (n,$\gamma$) reactions is 99.94\%.  

We conclude that the total UCN transmission of $92.6^{+1.2}_{-1.8}$\%
based on experimental results of the LPC Caen group accounts for all
effects leading to losses of UCN in the GS30 layer.

\subsection{ PSD cut efficiency and background rejection}\label{sec:bgreject}

Estimates of the signal efficiency and background rejection due to the
PSD versus $Q_L$ cut are estimated using an extended maximum
likelihood fit of the PDF templates described in
Section~\ref{sec:sim}.  The PDFs, binned in PSD and $Q_L$, are
labelled as $P_{i}( PSD, Q_L )$, where $i$=( 1n, Nn, 1$\gamma$,
N$\gamma$, 1n1$\gamma$, 1n deadtime, or 0n0$\gamma$).  The number of
each type of event is estimated as $N_i$ by minimizing a negative log
likelihood that is calculated as a sum over all $M$ of the PSD$^j$ and
$Q_L^j$ measurements in the data as:

\begin{equation}
-\ln{(L)} = \sum_{i}^{7} N_i - \sum_{j}^{M} \ln{ \sum_{i}^{7} P_i( PSD^j, Q_L^j) }.
\end{equation}

The data used were taken on West-1 for the time period from the start
of the three second proton beam.  Data were taken from all channels
for times 10~s to 280~s as shown in Fig.~\ref{fig:protonCycle}.  The
first three seconds of this data include a gamma flash and fast
neutrons.  Note that the single gamma background should include these
backgrounds as well, since they are taken over the sime time period
where the proton pulse has come, but the UCN gate valve was closed.

\begin{figure}[!htpb]
\centering 
\includegraphics[width = 0.45\textwidth] {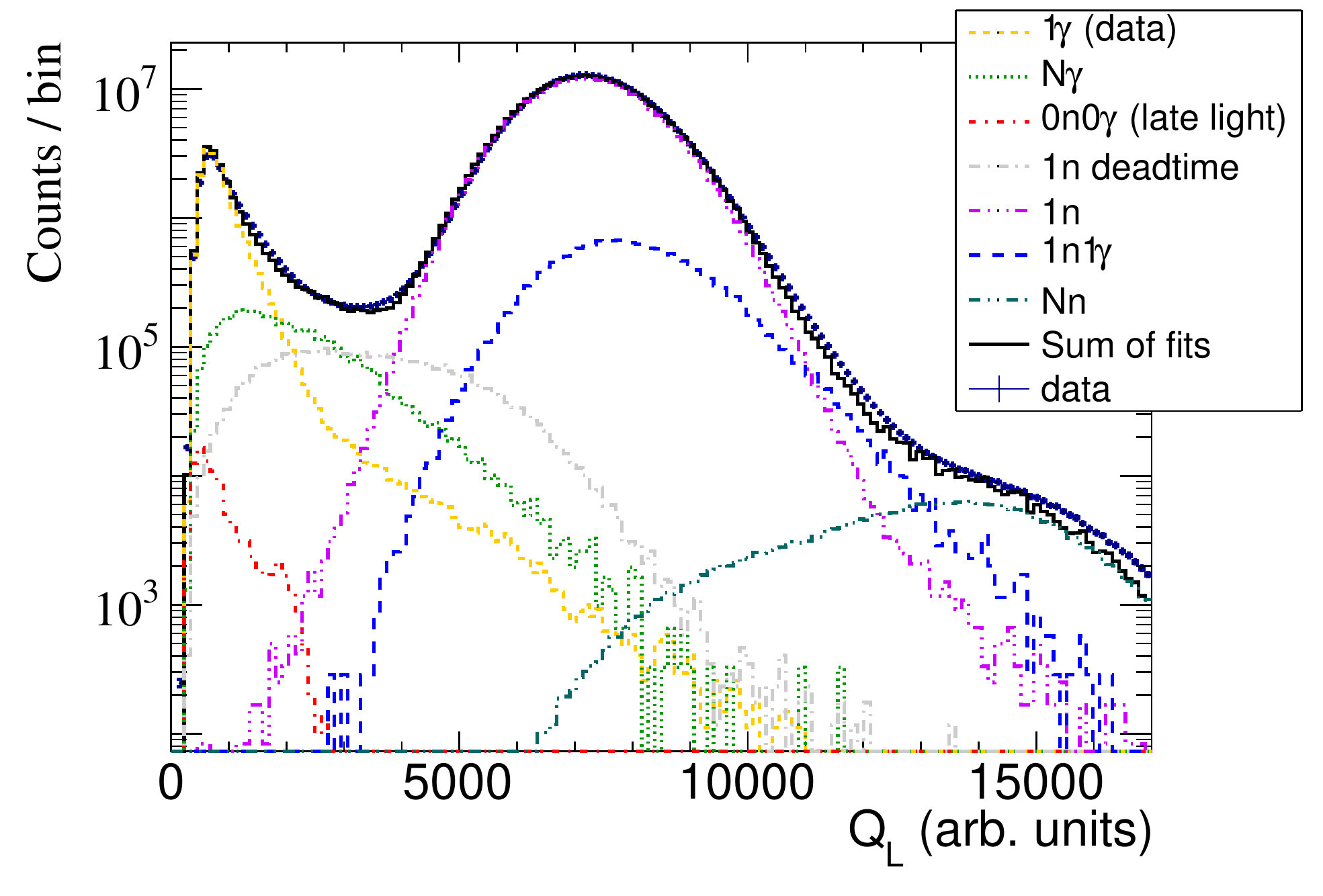} 
\includegraphics[width = 0.45\textwidth] {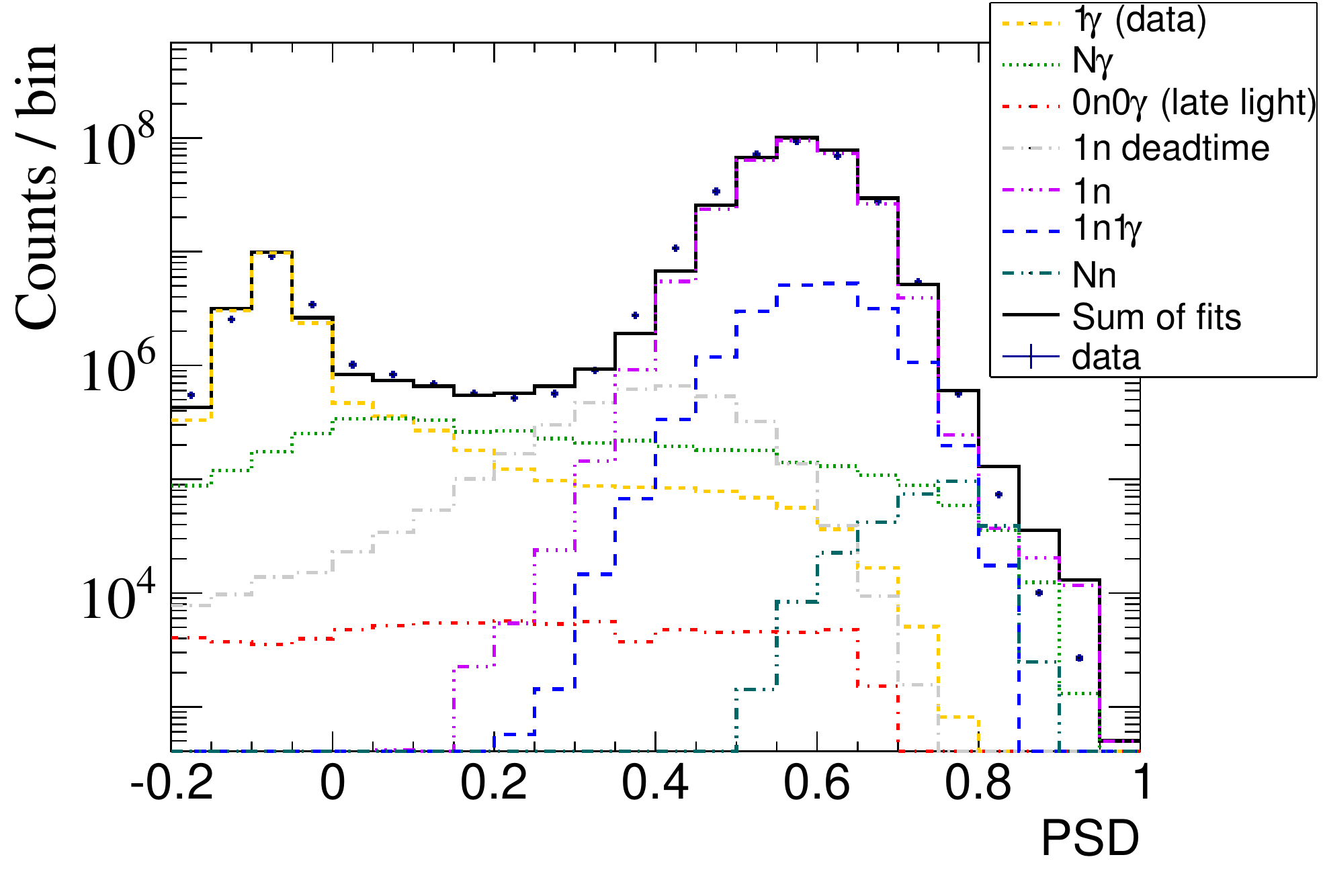}  
\caption{ Template fit results in the one dimensional projections
  along the total event charge (top) and along the PSD (bottom)
  (colour online).  These data are from the start of the main proton
  beam pulse, from 10 s to 280 s in the UCN cycle.}
\label{fig:fits}
\end{figure}

A projection of the fit results onto the PSD and $Q_L$ axes is shown
in Fig.~\ref{fig:fits}.  All of the features seen in the data are
reproduced in the fit, although the chi-squared per degree of freedom
(DOF) of the fit is rather poor ($\chi^2$/DOF~$= 775596/140$ in
$Q_{L}$ and $\chi^2$/DOF~$= 6.68\times10^8 / 20$ in PSD).  We
attribute the differences between the simulation and data to details
that are not properly modelled, such as any contribution from light
leaks, and PMT after-pulsing.  For the $Q_L$ distribution, differences
in gain between the nine channels of the detector may contribute.

\subsection{Cut detection efficiency and background rejection estimates}

Using the template fit, the neutron detection efficiency and
background contamination are computed for different cut values in PSD
and $Q_L$.  The PDFs representing neutron signals include single
neutron (1n), multiple neutrons (Nn), dead-time neutrons (1n
deadtime), and single neutron single gamma (1n1$\gamma$).  The
background rates are extracted from the single gamma (1$\gamma$),
late-light (0n0$\gamma$), and multiple gamma (N$\gamma$) templates.
If the total number of neutrons in the templates is $N_n$, and the
number of neutrons above a given cut value is $N_n^{cut}$, then the
neutron efficiency due to background rejection cuts is defined as:

\begin{equation}
\epsilon_n = \frac{N_n^{cut}}{N_n}.
\end{equation}

\noindent If the number of events in the background templates above a
given cut value is $N_{\gamma}^{cut}$, then the background
contamination fraction is defined as:

\begin{equation}
\eta_{\gamma}= \frac{N_{\gamma}^{cut}}{N_n},
\end{equation}

\noindent and the background rejection as:

\begin{equation}
\epsilon_{\gamma} = 1 - \eta_{\gamma}.
\end{equation}

\begin{figure}[!htpb]
\centering \includegraphics[width =0.45\textwidth, angle=0]{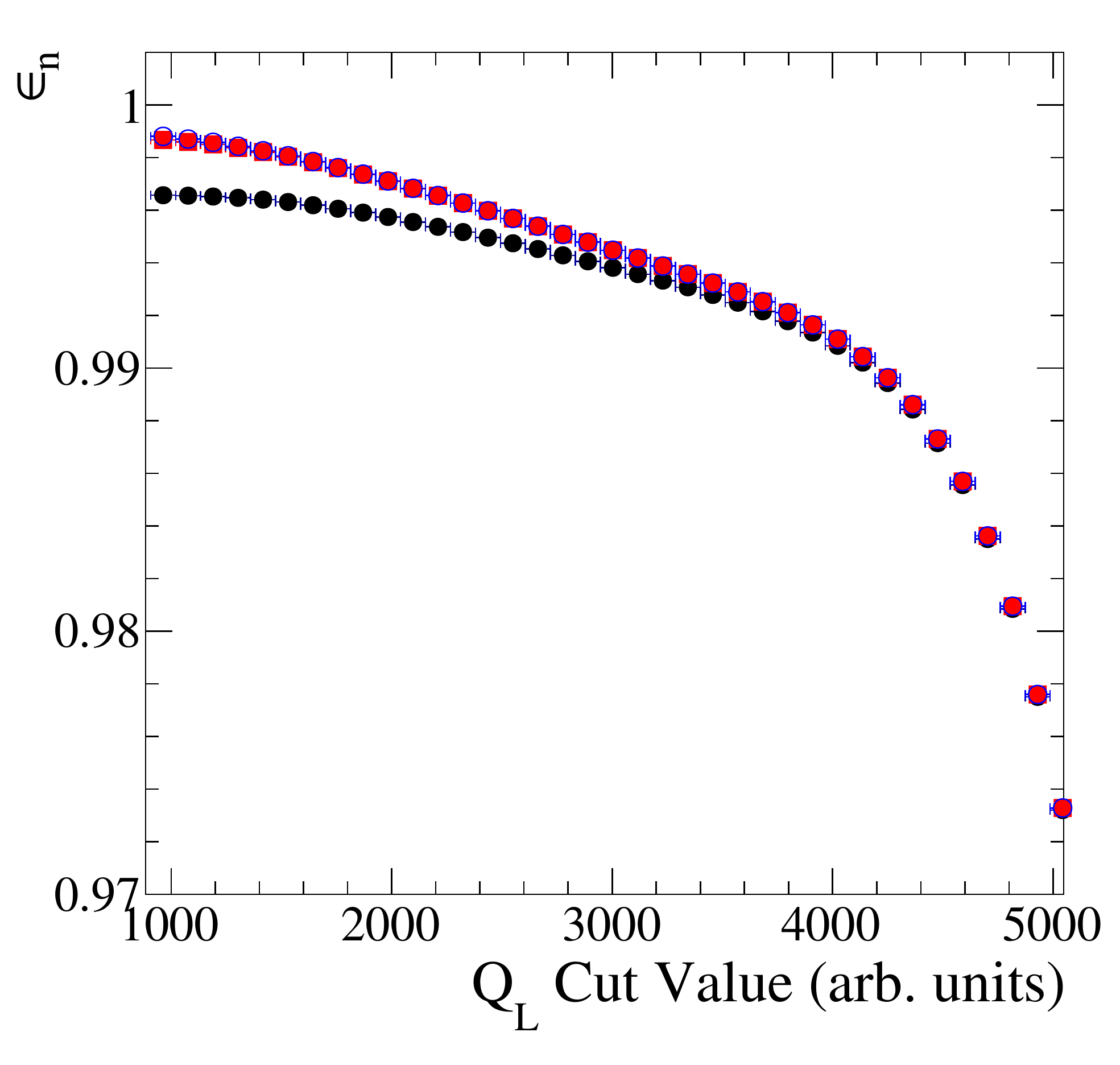}
\centering \includegraphics[width =0.45\textwidth, angle=0]{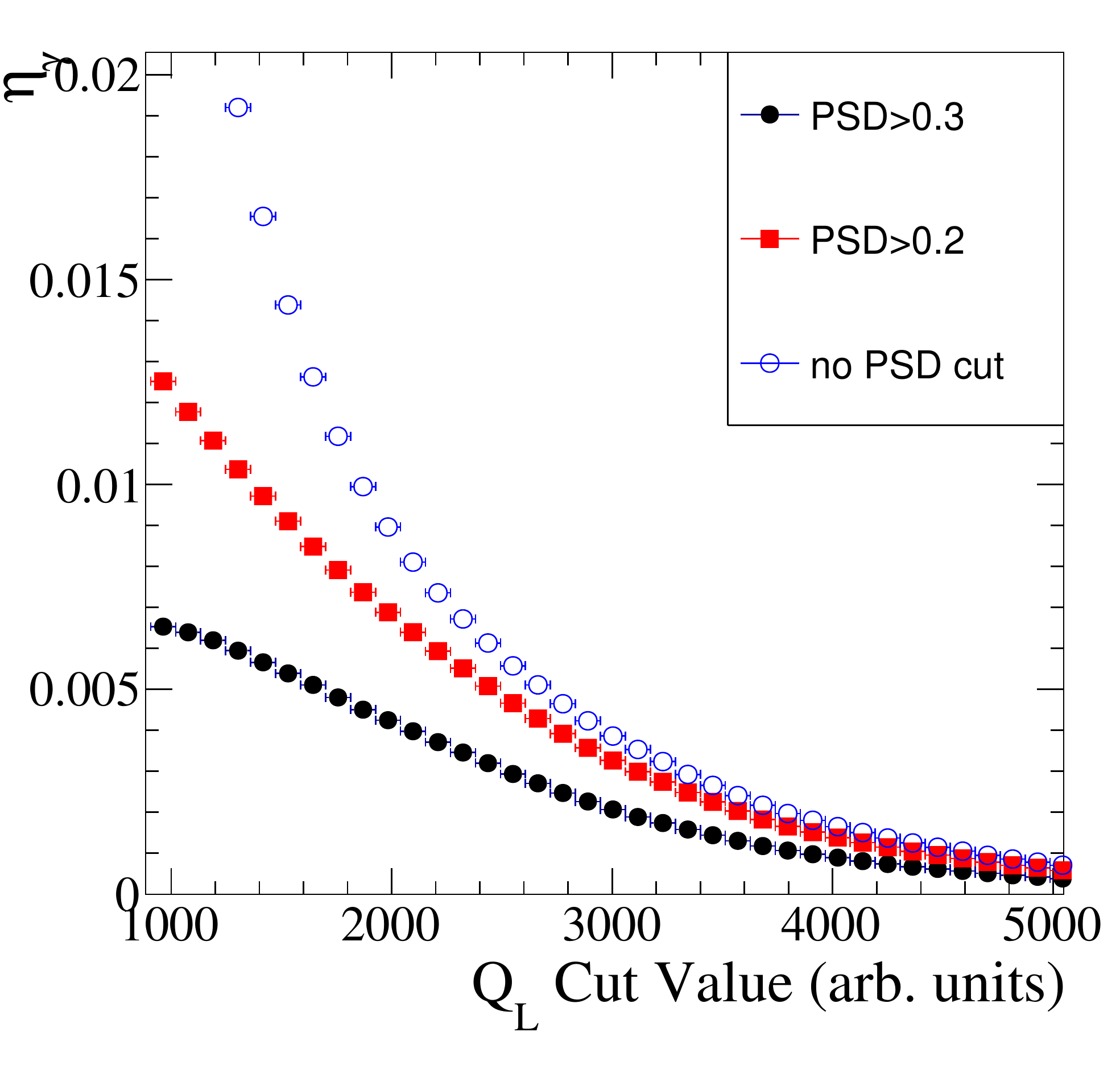}
\caption{ Neutron cut efficiency (top panel) and background
  contamination (bottom panel) for different cuts on $Q_L$.  In both
  panels the blue open circles have no PSD cut, the red squares are
  with cut on PSD$>0.2$, and the black circles are for a cut on
  PSD$>0.3$ (colour online). }
\label{fig:detectionEff}
\end{figure}

Figure~\ref{fig:detectionEff} shows the neutron efficiency and
background contamination during the entire UCN cycle for two PSD cut
values.  For higher cut values, the efficiency is slightly reduced,
but the background contamination is also reduced.


\begin{figure}[!htpb]
\centering \includegraphics[width = 0.45\textwidth, angle=0]{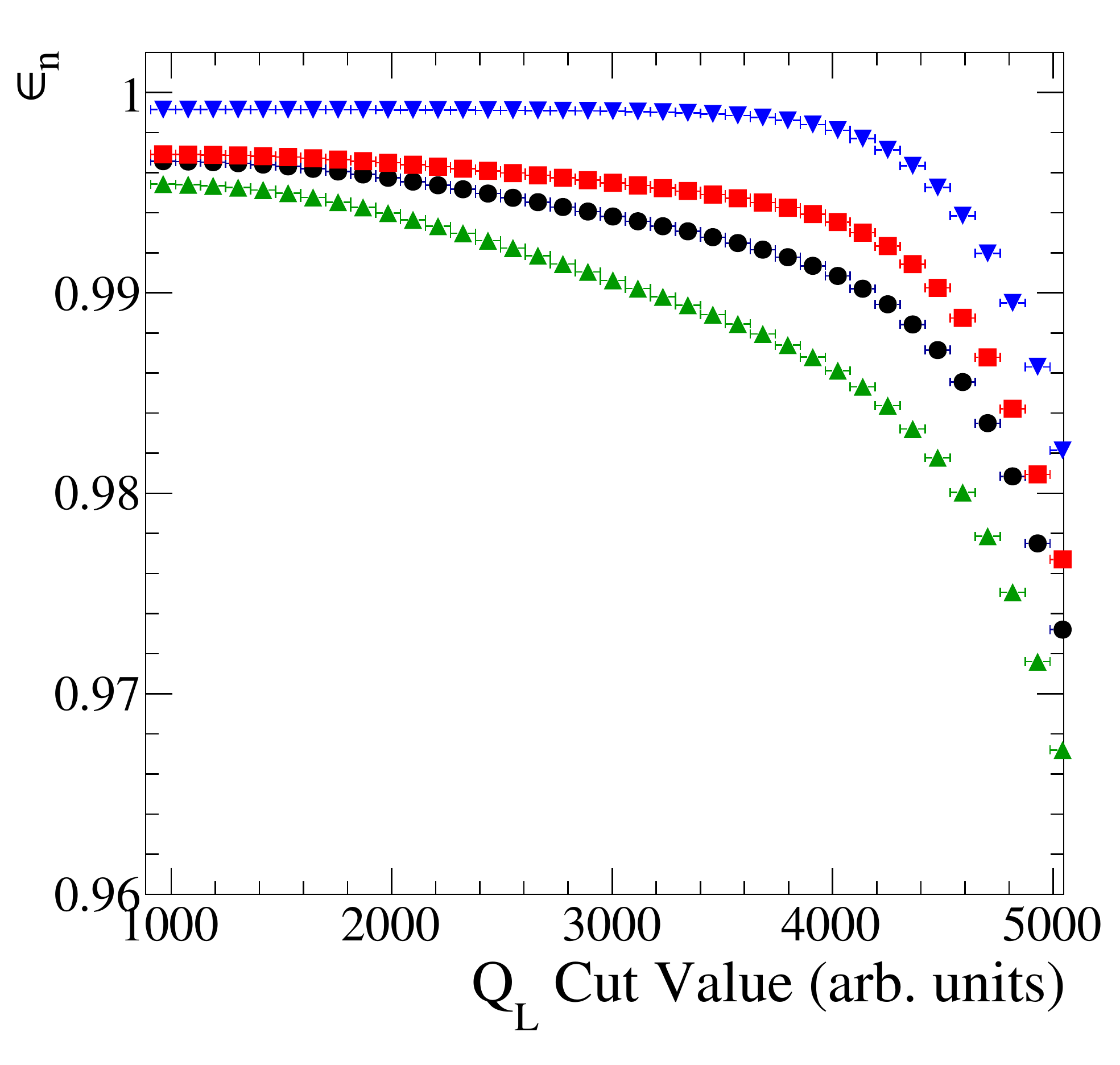}
\centering \includegraphics[width = 0.45\textwidth, angle=0]{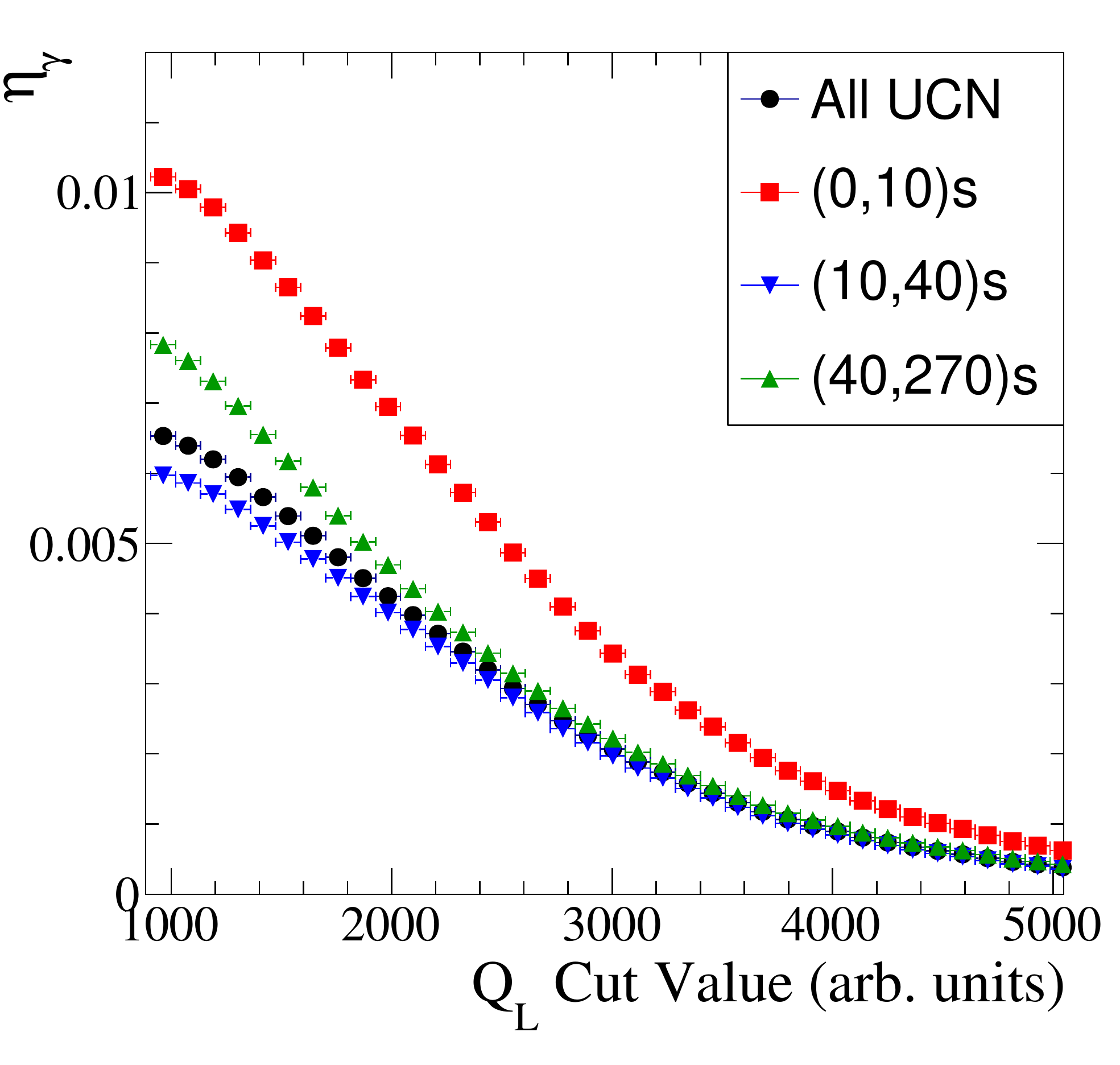}
\caption{ Neutron cut efficiency (top panel) and background
  contamination (bottom panel) for different times since the UCN
  production.  Shown in black are all UCN from 0~s to 270~s after the
  proton beam turns off.  The red boxes are for high rate over the
  first 10~s after the proton beam starts, the blue downward-triangles
  are for middle rate UCN over the next 30~s, and the green
  upward-triangles are for low rates over the last 230~s.  These plots
  have a cut PSD$>0.3$ (colour online).}
\label{fig:detectionEffLayered}
\end{figure}

The neutron and background rates vary over UCN cycle.  To study the
neutron detection efficiency and background contamination at different
rates on the West-1 beamline, the data was split into three time
periods during the UCN cycle: high rates of 100~kHz to 50~kHz at times
between 0~s an 10~s after the proton beam arrives, middle rates of
50~kHz to 20~kHz at times between 10~s and 40~s, and low rates of
20~kHz to 100~Hz beetween 40 and 270~s.  Each of these data sets was fit
using the template fit and then the efficiency was calculated with cut
on PSD$>0.3$ using the parameters from each fit.  As shown in
Fig.~\ref{fig:detectionEffLayered}, the high rate data had more
background contamination than the other rates, due to the larger
fraction of events with pile-up effects, and due to the proton beam
being on for the first three seconds.  The lowest rate data has a
higher contamination than the mid-rate events due to the higher ratio
of background to signal events.  Using a cut on $Q_L > 3000$~ADC and
PSD$>0.3$, the neutron efficiency was $99.5\pm0.5$\% and the
background contamination was $0.3\pm0.1$\%.

The $^6$Li detector therefore has a very good background rejection due
to the signal shape variation between light-guide background (Cerenkov
events) and the scintillation events from the lithium glass.

\subsection{Overall Detector Efficiency Estimate}

The overall detection efficiency, for the UCN rates and energies
available at the West-2 beamport, including effects of GS30
transmission, effective area, and background rejection, is
$89.7^{+1.3}_{-1.9}$\%, which is dominated by the uncertainty in the
absorption in the GS30 layer.  The transmission measurement of the
GS30 was performed by the CAEN group at ILL, with a different UCN
energy spectrum, and so there remains the caveat that this efficiency
is for UCN energies above the Fermi potential of the $^{6}$Li glass.
The effects of surface imperfections is not included in this estimate,
and is a possibly the source of the differences between the relative
rates observed between the different channels of the detectors when
compared to the area they present to the UCN beamline.

The rate dependent portion of the uncertainty, which is most
concerning to measurements requiring a stable rate, could be improved
by applying the statistics of random signals and backgrounds to model
the expected rates.  This would represent an improvement to the simple
fit with unconstrained fractions of different types of pile-up
described in this paper.

\section{ Relative rate comparison }\label{sec:relative}

\subsection{ Comparison to Cascade detector }

\begin{figure}[!htpb] 
\centering \includegraphics[width=0.45\textwidth]{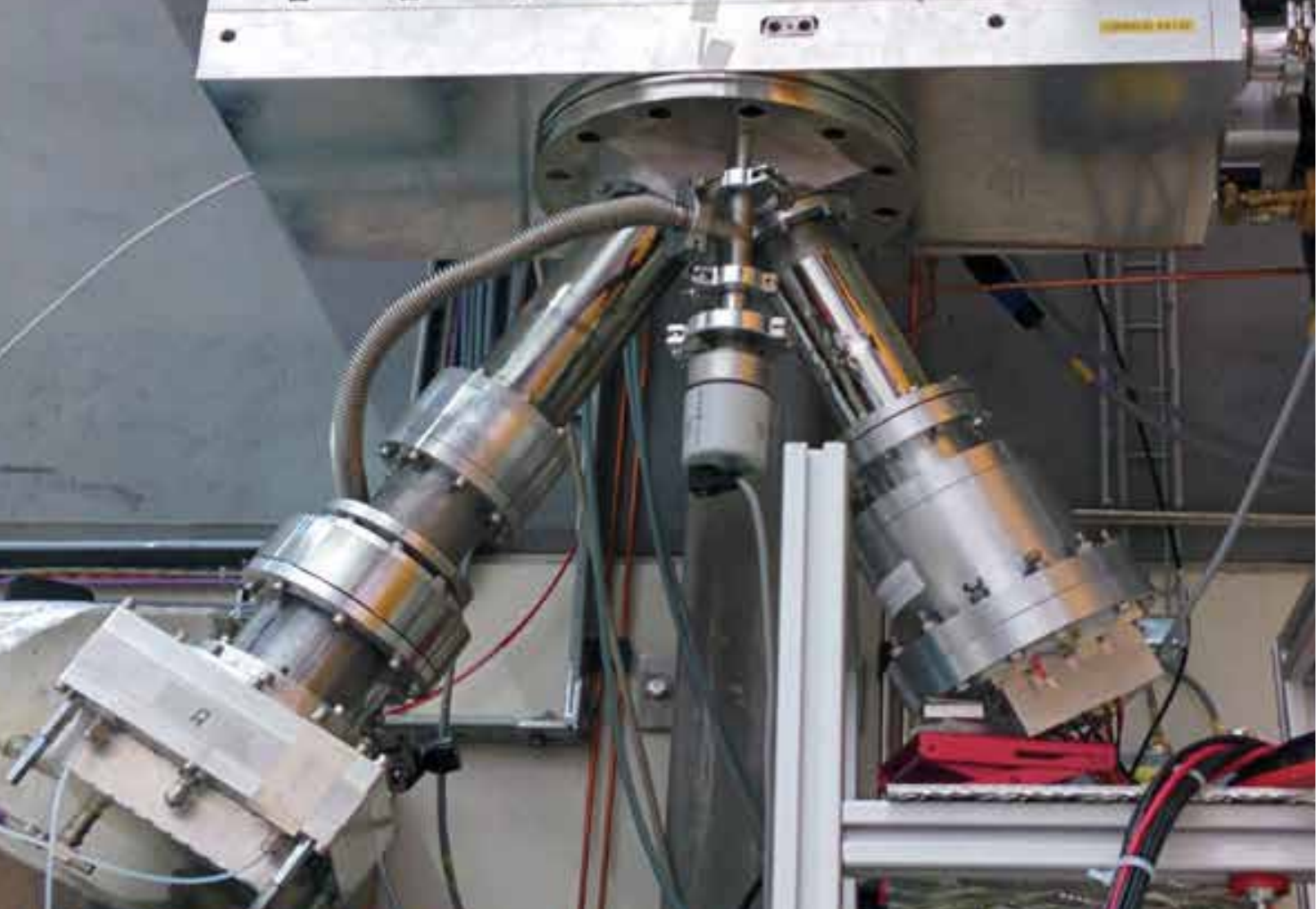}
\caption{Configuration for splitting the UCN into two detectors on the
  West-2 beamline.  The Cascade detector is on the left and the $^6$Li
  detector is on the right (colour online).}
\label{fig:yConfig}
\end{figure}

A y-shaped UCN beam splitter to divide the UCN evenly into two ports
was used on the West-2 beamline at PSI to compare the rates of UCN
detected by a Cascade detector and our $^6$Li detector.  The detectors
in this y-configuration are shown in Fig.~\ref{fig:yConfig}.

The connection of the $^6$Li detector to the y-shaped beam splitter
was a 20~cm long NiMo coated guide with a diameter of 75~mm.  The
Cascade detector has a 110~mm diameter aperture, and two adapter
flanges were needed to go from 110~mm to 70~mm to 75~mm to connect to
the other side of the y-configuration.  The Cascade detector therefore
sits $\sim20$~cm lower than the lithium glass detector and has a
longer path for the UCN to pass.

\subsection{Cascade Detector}

The Cascade UCN detector is a GEM-based neutron detector with a single
200~nm thick layer of $^{10}$B deposited on a 100~$\mu$m thick
aluminum entrance foil. The boron captures neutrons and releases an
$\alpha$ and $^7$Li particle,

\begin{equation}
^{10}B + n \rightarrow \alpha + {^7}Li.
\end{equation}

An Ar/CO$_2$ mixture is used as a detection gas.  Due to the low Z
materials, this detector picks up negligible $\gamma$ background. The
employed detector has a $10 \times 10$~cm$^2$ square shaped sensitive
area which is divided into 64 individually read out pixels.  The
detector comes with its own proprietary data acquisition system.  The
data acquisition is based on a complex pattern recognition algorithm
which is performed online in the detector's FPGA electronics.  This
allows the software to define what patterns are accepted as a neutron
event.


\subsection{Overview of the Measurement Method}

Both detectors were placed in the y-configuration at the West-2 port
to allow for a comparison of the averaged UCN rate over the course of
the beam cycle for both detectors.  The timing calibration between the
two detectors' Data AcQuisition (DAQ) was done at the few second level
by comparing the time reported by the DAQ computers.  A closer
matching of the time in the analysis of the data was performed at
about the 0.05~s level by aligning the times when the UCN rate was
increasing when the main proton beam arrives, corresponding to times
around 10 s in Fig.~\ref{fig:protonCycle}.

\begin{figure}[!htpb]
\centering \includegraphics[width=0.45\textwidth]{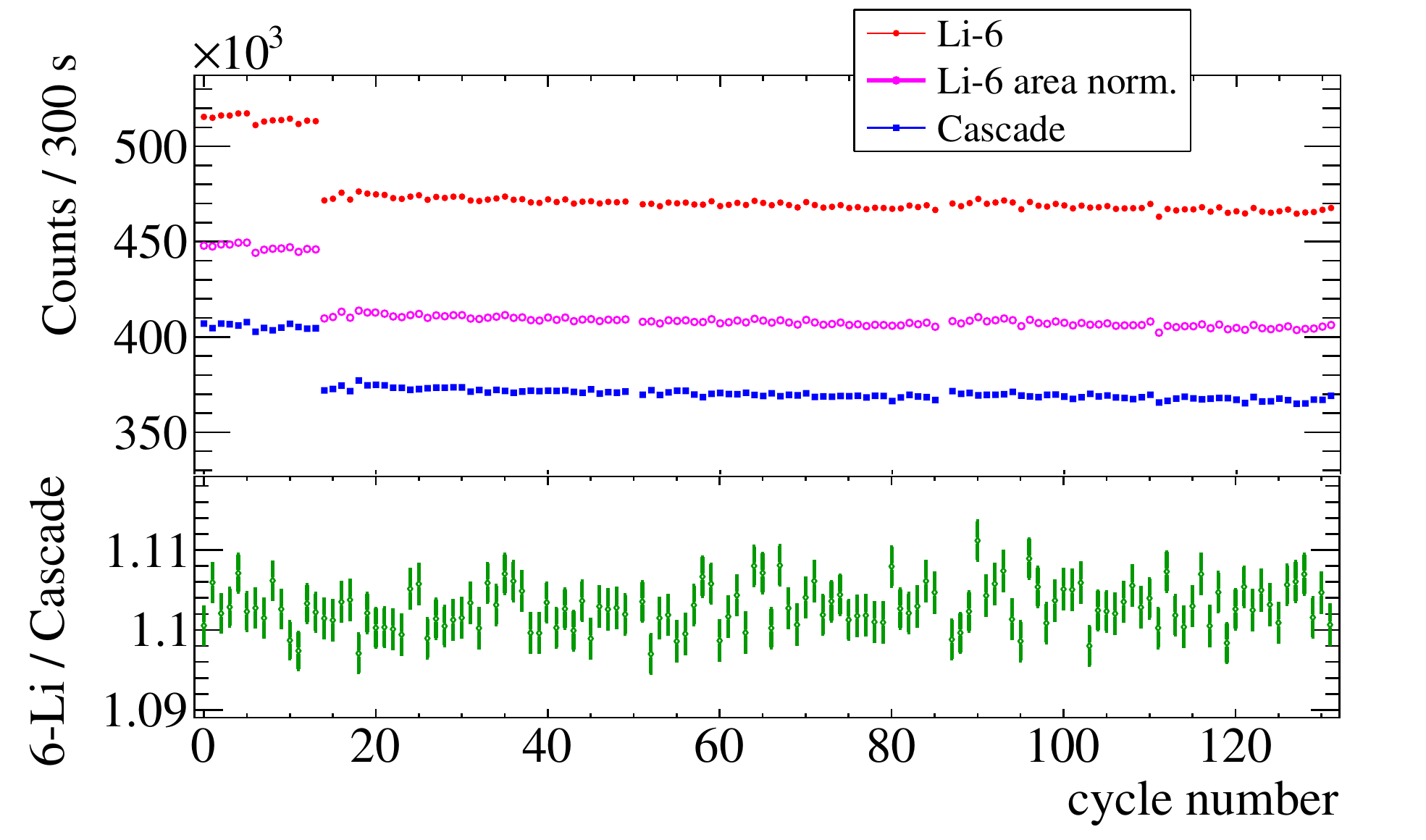}
\caption{Detector count per UCN cycle is shown in the top panel for
  the ${^6}$Li detector by red filled circles after cuts on $Q_L>3000$
  and PSD>0.3, the Cascade detector by blue filled squares, and the
  area normalized ${^6}$Li count by magenta open circles.  The bottom
  panel shows the ratio of the area normalized count in the $^{6}$Li
  detector to the Cascade detector (colour online).}
\label{fig:ratecompare}
\end{figure}

The top panel of Fig.~\ref{fig:ratecompare} shows UCN counts detected
by both detectors in each cycle.  The $^{6}$Li detector counts are
after cuts on $Q_L>3000$ and PSD>0.3 were applied.  We observe that
during the course of the measurement, the UCN count from the PSI
source was changing, and that both detectors track this change in the
same way.  The smallest aperature that the UCN must pass to get to the
detectors is used as a worst case scaling of the counts seen by the
6-Li detector.  The scaling is used to calculate a normalized count to
compare with the Cascade detector.  The ratio of the areas used in the
scaling is $(70$~mm$/ 75$~mm$)^2 = 0.871$.  The bottom panel of
Fig.~\ref{fig:ratecompare} shows a ratio of the area normalized count
in the $^{6}$Li detector to the count in the Cascade detector.  This
comparison shows that for the spectrum of UCN in this beamline the
${6}$Li detector had at least $10.279 \pm 0.024$(stat)\% more counts
than the Cascade detector.

The additional flanges required to adapt the Cascade detector to the
$75$~mm diameter y-configuration introduce a large uncertainty in
measuring the Cascade counting rate.  A GEANT4 UCN simulation of the
y-configuration was prepared, using NiMo Fermi potential of 235~neV
for the beampipes and adapter flanges.  The number of UCN reaching
each detector was found to strongly depend on the assumed initial UCN
spectrum reaching the entrance of the y-configuration ports.  We
therefore conclude that the $^{6}$Li detector is at least as efficient
as the Cascade detector, but that the difference in UCN counts that we
observed could entirely be due to the additional flanges required for
the Cascade detector.


\begin{figure}[!htpb]
\centering \includegraphics[width = 0.45\textwidth]{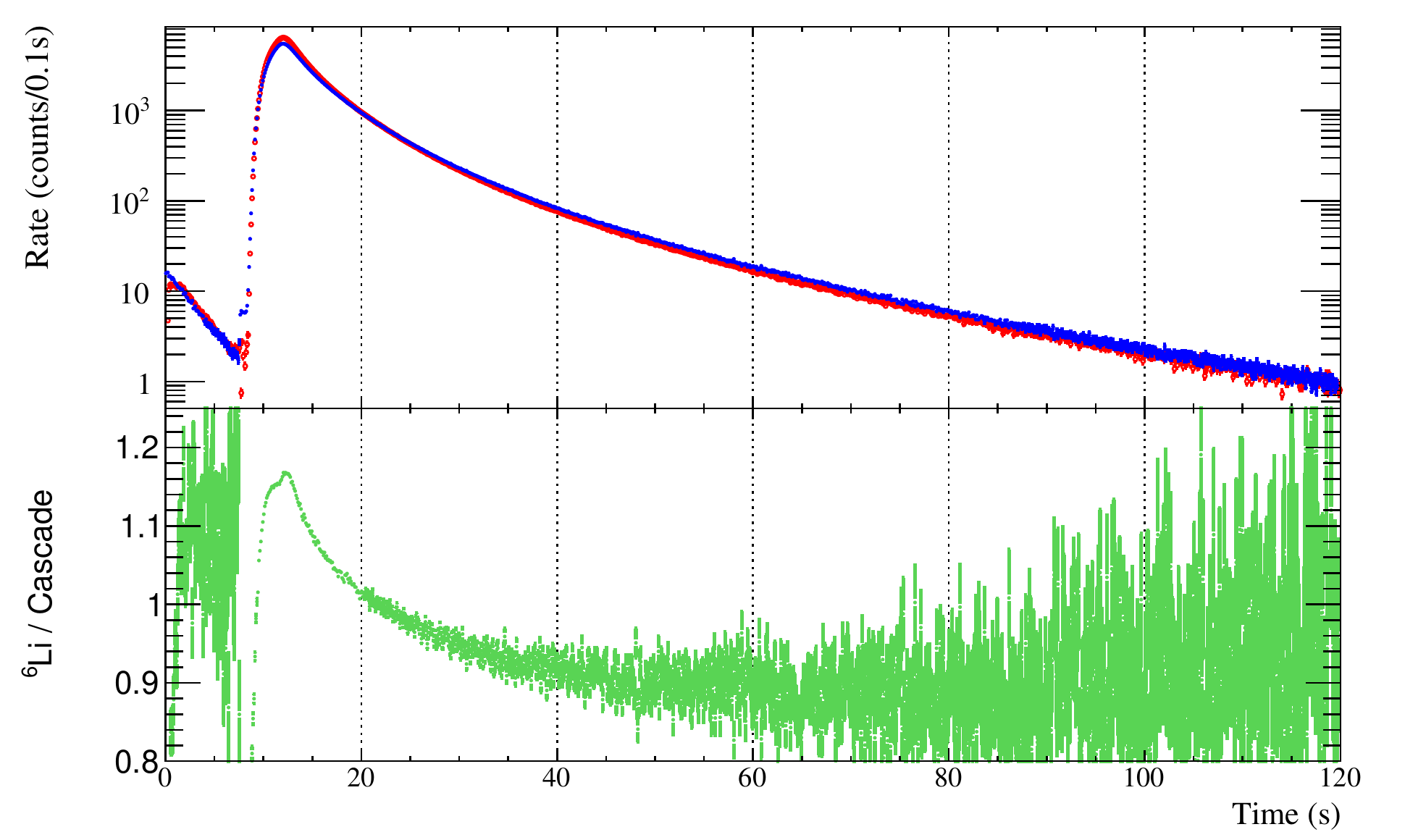}
\caption{ The top panel shows the counts per 0.1~s bin averaged over
  130 UCN cycles for the area normalized $^6$Li detector (red) and the
  Cascade detector (blue).  The bottom panel shows the ratio of counts
  in the $^6$Li detector to the Cascade detector (colour online). }
\label{fig:averagedRate}
\end{figure}

The UCN detection rate in 0.1 second bins since the beginning of the
300~s UCN cycle averaged over 130 beam cycles is compared in
Fig.~\ref{fig:averagedRate}.  Again the $^{6}$Li detector counts are
after cuts on $Q_L>3000$ and PSD>0.3.  This comparison includes the
area normalization and shows that $^{6}$Li detector detect more UCN at
higher rates near the beginning of the UCN production than the Cascade
detector.  During the course of the UCN cycle the energy spectrum of
the UCN reaching the detectors changes: the faster UCN reach the
detectors sooner, and later in the beam cycle, the slower UCN reach
the detectors.  The $^{6}$Li with effective Fermi potential of
103.4~neV does not detect the slowest UCN, making the Cascade detector
more efficient for lower energy UCN (down to the 55~neV of Al).

We conclude that the $^{6}$Li detector has a detection efficiency at
least as large as the Cascade detector for UCN with energies above the
Fermi potential is 103.4~neV.

\section{Conclusion}

In this paper we have presented a $^{6}$Li-based fast scintillation
counter and a detailed electronics simulation which has been used to
estimate the detector efficiency and background rejection for the data
collected at the PSI UCN source.  The absolute detector efficiency is
found to be $89.7^{+1.3}_{-1.9}$\%, with a background contamination of
$0.3\pm0.1$\%.  Using comparisons of UCN rates from the different
tiles of the $^{6}$Li detector, we have demonstrated that the detector
is stable at the 0.06\% level or better, and that the variation in
efficiency between the detector tiles is less than 5\%.  Finally we
have shown that the $^{6}$Li detector is at least as efficient as the
Cascade detector for UCN with energies above the Fermi potential of
the lithium glass.  Counting rate differences observed between the two
detectors can be entirely explained by the additional flanges needed
to connect the detectors to the same UCN beamline.

\section{Acknowledgements}

We acknowledge support of the National Science and Engineering
Research Council (NSERC) and Canadian Foundation for Innovation (CFI)
in Canada.  We thank our colleagues at PSI for allowing us to conduct
the detector tests at their facility.  We thank staff and colleagues
at TRIUMF who have provided support and feedback in the design and
testing of the detector.  We are grateful to Andrew Pankywycz at the
University of Manitoba who machined the light-guides and aluminium
detector housing.  Finally we thank University of Winnipeg technician
David Ostapchuck who helped with various aspects of the design and
construction of the detector.


\bibliography{bmc_article}

\end{document}